\documentclass[12pt]{article}

\usepackage{geometry}
\usepackage{amsmath, amssymb, mathabx, comment, mathrsfs}
\usepackage{graphicx, color}
\usepackage{array, multirow}
\usepackage[font=small,labelfont=bf]{caption} 
\usepackage{lipsum}
\usepackage{amsfonts}
\usepackage{tabularx}
\usepackage{booktabs}
\usepackage{graphicx}
\usepackage{epstopdf}
\usepackage{algorithmic}
\usepackage{mathtools}
\usepackage{tikz}
\usepackage{todonotes}
\definecolor{darkgreen}{rgb}{0, .5, 0}

\ifpdf
\DeclareGraphicsExtensions{.eps,.pdf,.png,.jpg}
\else
\DeclareGraphicsExtensions{.eps}
\fi

\AtBeginDocument{%
\setlength{\oddsidemargin}{\dimexpr(\paperwidth-\textwidth)/2-1in}%
\setlength{\evensidemargin}{\oddsidemargin}%
\setlength{\topmargin}{%
\dimexpr(\paperheight-\textheight)/2-\headheight-\headsep-1in}%
}

\usepackage[shortlabels]{enumitem}

\usepackage[sort]{natbib}

\setcitestyle{numbers,square,comma}
\usepackage{hyperref}
\hypersetup{colorlinks=true,
urlcolor=blue,
linkcolor=blue,
citecolor=blue,
bookmarksdepth=paragraph}
\usepackage[nameinlink]{cleveref}
\crefname{equation}{}{}
\crefname{section}{section}{sections}
\crefname{figure}{figure}{figures}
\crefname{table}{table}{tables}
\crefname{example}{example}{examples}
\crefname{proposition}{proposition}{propositions}
\Crefname{section}{Section}{Sections}
\Crefname{figure}{Figure}{Figures}
\Crefname{table}{Table}{Tables}
\Crefname{definition}{Definition}{Definitions}
\Crefname{theorem}{Theorem}{Theorems}
\Crefname{remark}{Remark}{Remarks}
\Crefname{example}{Example}{Examples}
\Crefname{proposition}{Proposition}{Propositions}
\numberwithin{equation}{section}

\usepackage{amsfonts,amsmath,amssymb,amsthm}

\newtheorem{theorem}{Theorem}[section]
\newtheorem{lemma}[theorem]{Lemma}
\newtheorem{corollary}[theorem]{Corollary}
\newtheorem{proposition}[theorem]{Proposition}
\theoremstyle{definition}
\newtheorem{definition}[theorem]{Definition}
\newtheorem{remark}[theorem]{Remark}

\newtheorem{example}[theorem]{Example}

\newcommand{\R}{\mathbb{R}}
\newcommand{\N}{\mathbb{N}}

\NewDocumentCommand{\E}{o}{%
    \mathbb{E}^g\IfValueT{#1}{\big( {#1} \big)}
}
    \NewDocumentCommand{\Eg}{o}{%
        \mathbb{E}^g\IfValueT{#1}{\big( {#1} \big)}
    }

\renewcommand{\N}{\mathbb{N}}
\renewcommand{\P}{\mathbb{P}}
\renewcommand{\R}{\mathbb{R}}

\NewDocumentCommand{\Flow}{o}{%
    \operatorname{Flow}{\IfValueT{#1}{\left({#1}\right)}}
}
\NewDocumentCommand{\supp}{o}{%
    \operatorname{supp}{\IfValueT{#1}{\left({#1}\right)}}
}

\newcommand{\eqdef}{\ensuremath{\stackrel{\mbox{\upshape\tiny def.}}{=}}}

\newcommand{\calF}{\mathcal{F}}

\usepackage[normalem]{ulem}
\usepackage{booktabs}
\usepackage{float}

\usepackage[skins,breakable]{tcolorbox}
\usepackage{graphicx}
\usepackage{subcaption}
\definecolor{extensionteal}{HTML}{99E0D9}
\definecolor{extensiongrey}{HTML}{F2F2F2}

\allowdisplaybreaks[2]

\title{Reconstructing risk preferences from data}

\usepackage{hyperref}

\author{%
    Anastasis Kratsios\thanks{Corresponding author.}\,\,\thanks{Department of Mathematics and Statistics, McMaster University; Vector Institute for Artificial Intelligence, Canada. \href{mailto:kratsioa@mcmaster.ca}{\texttt{kratsioa@mcmaster.ca}}.}
    \and
    Silvana M. Pesenti\thanks{Department of Statistical Sciences, University of Toronto, Canada. \href{mailto:silvana.pesenti@utoronto.ca}{\texttt{silvana.pesenti@utoronto.ca}}.}
    \and
    Philipp Zimmermann\thanks{Institute of Mathematics, \'{E}cole Polytechnique F\'{e}d\'{e}rale de Lausanne (EPFL), Switzerland. \href{mailto:philipp.zimmermann@epfl.ch}{\texttt{philipp.zimmermann@epfl.ch}}.}
}
\date{\today}

\usepackage{xcolor}
\definecolor{cobalt}{rgb}{0.0, 0.28, 0.67}

\begin{document}

\maketitle

\begin{abstract}
We consider the problem of reconstructing risk preferences from data. Specifically, for a finite dataset consisting of contingent claims and their evaluations, we aim at reconstructing a risk measure that is finite-valued, cash-additive, convex, and Lipschitz continuous. Our first result establishes necessary and sufficient conditions for a dataset to be compatible with a risk measure, that is the evaluations are stemming from a risk measure applied to the contingent claims. Our second result provides a reconstruction formula for compatible datasets, which we term \emph{universal risk measure}. We further extend the reconstruction to monotone risk measures and establish sharp Lipschitz constants for both general and monotone reconstructions. We further show that we explicitly reconstruct linear preferences and evaluations by the Expected Shortfall. We conclude with a numerical application to European options. 
\end{abstract}

\noindent
\textbf{Keywords:} Risk Elicitation, Convex Risk Measures, Robo-Advising, Inverse-Problems.

\section{Introduction}
\label{s:Intro}

{Our starting point is a version of the classical preference elicitation problem arising in, e.g., robo-advising \cite{capponi2022MS}. Suppose that a user's preferences are modeled by a linear pricing / valuation functional. The user is then queried with contingent claims $X_1,\ldots,X_N$ and reports their evaluation
\begin{equation*}
        y_n
    =
        \mathbb{E}[Z^\star X_n]\,,
        \quad n=1,\ldots,N\,,
\end{equation*}
where $Z^\star$ is a pricing (Radon--Nikodym) density. As we only observe $\{(X_n, y_n)\}_{n = 1}^N$ and not the pricing functional itself, we aim at inferring the user's valuation functional from finitely many queries.  Figure~\ref{fig:OurPaperTLDR} illustrates this robo-advising setting. In a first phase, the robo-advisor gathers informative data on the user's preferences.  In the second phase, which is the object of study in this paper, the robo-advisor aims at building a maximally regular risk measure consistent with the user's quoted data. 
\begin{figure}[H]
    \centering
    \includegraphics[width=0.75\linewidth]{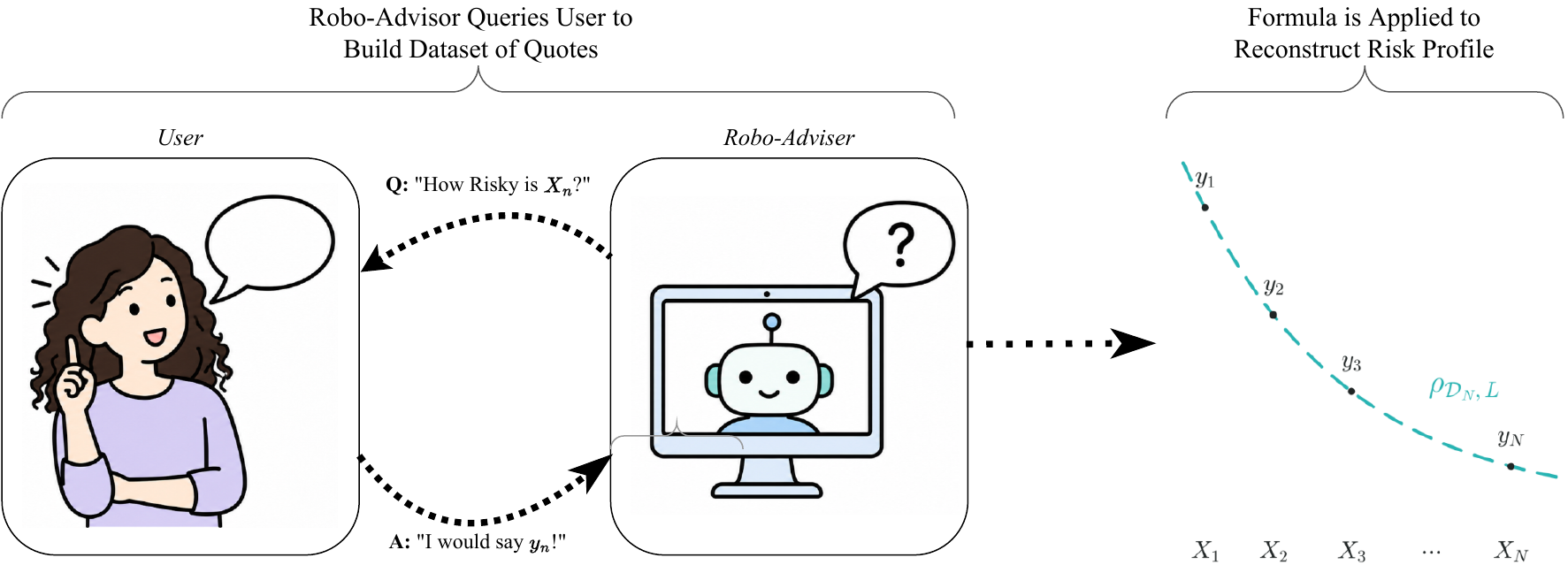}
    \caption{The robo-adviser provides the user with a sequence of queries $X_1, \ldots, X_N$ and the user assesses their risk $y_1, \ldots, y_N$. The robo-adviser then aims at building a risk measure consistent with the user's quoted data.  
    }
    \label{fig:OurPaperTLDR}
\end{figure}
In many practical applications, users evaluate contingent claims through nonlinear measures of risk. Motivated by this observation, we replace the linear pricing rule by an unknown risk measure and ask whether finitely many quoted evaluations determine a globally defined risk measure $\rho$ compatible with the observed data, that is if there exists a risk measure satisfying $y_n = \rho(X_n)$ for all $n = 1, \ldots, N$. Note that here, we assume that the contingent claims and the evaluations are given and do not study the problem of how to design the contingent claims.

Elicitation of utilities and risk preferences has a long history in the economic and risk management literature; indicatively see \cite{wakker1996eliciting, abdellaoui2000parameter, bleichrodt2000parameter}, and auctions in algorithmic game theory; cf.~\cite{MennleSeuken2020Axiomatic,MennleSeuken2021}. Recently, it has entered the field of reinforcement learning (RL), cf.~\cite{cheng2023eliciting} and references therein, who design interactive questionnaires to elicit risk aversion for spectral risk measures. 
Different, but closely related, is the identification of preferences in combinatorial auctions and market design, see e.g., \cite{BlumJacksonSandholmZinkevich2004,LahaieParkes2004,SandholmBoutilier2006}, and another stream of works considers robust decision-making under preference uncertainty, see e.g., \cite{guo2022MP,wang2023preference,bernard2026preference} for applications to risk management.

Typically in the preference elicitation literature, the user's preferences are modelled via a parametric family and the elicitation pertains to the user's risk aversion.   However, in this work, we do not restrict to a parametric family of risk measures. Rather we start with a dataset $\{(X_n, y_n)\}_{n = 1}^N$, consisting of risks $X_n $ (random variables) and risk evaluations $y_n$ (real numbers). In our analysis the dataset is fixed and we do not address the design of the queries $X_n$; we refer to \cite{MennleSeuken2021,cheng2023eliciting} for interested readers on interactive questioning. We first investigate which datasets are compatible with a risk measure, that need not be the user's risk measure. That is we answer the question which datasets can be seen as generated by a (latent) risk measure in that $y_n = \rho(X_n)$ for all $n = 1, \ldots, N$ for some risk measure $\rho$.  
Our analysis is a Whitney-type extension problem, that asks when finitely specified data admit a globally defined extension satisfying prescribed regularity or structural constraints. Classical instances of Whitney-type extension problems include Lipschitz extension theorems~\cite{mcshane1934extension,ok2025order}, continuous extension theorems~\cite{dugundji1951extension}, $C^{k,\alpha}$-extension problems~\cite{whitney1934analytic,
brudnyi1994generalizations,
fefferman2009whitney}, computationally efficient extension problems \cite{kratsios2026computationallyefficientclosedformcs1extensionformula}, and
Sobolev-Besov extension theory~\cite{jonsson1978whitney}. Here, the data consist of finitely many pairs of risks and their evaluations, and the structural constraints are risk measures that are cash-invariant and convex. The compatibility of a dataset with a risk measure is a Whitney-type question, that we address in Theorem~\ref{thrm: risk fields characterize}. Specifically, we provide an answer via the notation of risk fields, which play the role of Whitney fields, namely, they constitute the finite compatibility conditions for global extension \cite{whitney1934analytic,fefferman2005sharp}.
Second, for compatible datasets, we propose an explicit reconstruction of a latent risk measure -- which we term \textit{universal risk measure}. Universal risk measures are constructed solely from a compatible dataset with the constraint that the risk evaluations $y_n$ equal the universal risk measure applied to the contingent claims $X_n$. We show that universal risk measures are cash-additive, convex, and Lipschitz continuous. We further investigate the case when users' preferences are monotone and modify the universal risk measure construction to preserve monotonicity. We also prove the optimal Lipschitz constant for universal risk measures and show how the universal risk measure reconstructs linear and Expected Shortfall preferences.

A related stream of literature is structure-preserving machine learning. In personalized portfolio allocation, for instance, a robo-advisor infers a user's personalized portfolio preferences by asking questions regarding their risk aptitude. While deep learning methodologies are adequate at interpolating between paired quotes, they are less suited to infer the user's risk profile. Our approach is thus to directly hard-code the problem structure into the candidate model architecture. This philosophy underlies a growing body of work on structure-preserving machine learning in mathematical finance. Examples include the universal arbitrage-free volatility surface models ~\cite{buehler2026sanos}, SDE-based approaches~\cite{cohen2023arbitrage}, universal causal kernels developed in~\cite{AcciaioPammerAnnie_2024} and their non-degenerate conditionally Gaussian versions~\cite{arabpour2026low}.

The paper is organised as follows.
Section~\ref{s:Prelim} characterises compatible datasets and introduces universal risk measures, which reconstruct risk preferences of compatible datasets. Section~\ref{sec:optimal-Lipschitz} develops the optimal Lipschitz constant of universal risk measures and Section~\ref{sec:linear-risk-ES} gives the explicit reconstruction of linear risk profiles and the Expected Shortfall. In Section~\ref{sec:numerical} we illustrate the reconstruction on a risk management example. All proofs are delegated to Appendix~\ref{s:Proofs} and Appendix~\ref{s:ToyExperiments__ss:Computation} contains additional information of the implementation.

\section{Reconstructing Risk Measures from Data}
\label{s:Prelim}

\subsection{Notation and Preliminaries}
We write $\mathbb{N}$ for the set of non-negative integers and $\mathbb{N}_+\eqdef \{n\in \mathbb{N}:\,n>0\}$ for the set of positive integers.  
For every $N\in \mathbb{N}_+$ we denote the ordinal $[N]\eqdef \{n\in \mathbb{N}:\, n\le N\}$ and $[N]_+\eqdef \{n\in [N]:\, n>0\}$. We further denote the standard $N$-simplex by
\[
\Delta_N
 \eqdef 
\left\{
\lambda=(\lambda_1,\ldots,\lambda_N)\in[0,\infty)^N:
\sum_{n=1}^N\lambda_n=1
\right\}
.
\]
For any subset $A$ of a vector space, we denote by $\operatorname{span}(A)$ the linear span of $A$ and by $\operatorname{co}(A)$ its convex hull. In particular, we may write
\[
\operatorname{co}\{x_1,\ldots,x_N\}
=
\left\{
\sum_{n=1}^N \lambda_n x_n :
\lambda\in\Delta_N
\right\}.
\]
If the vector space is endowed with a topology, then $\operatorname{int}(A)$ stands for its topological interior.

Given any lower semicontinuous (lsc), convex function $\varrho:H\to (-\infty,\infty]$ defined over a Hilbert space $H$, we denote its domain by $\operatorname{dom}(\varrho)\eqdef \{x\in H:\, \varrho(x)<\infty\}$ and we call $\varrho$ proper if $\operatorname{dom}(\varrho)\neq \emptyset$ and $\varrho(x)> - \infty$ for all $x \in H$.  When contextually clear, we write $\|\cdot\|$ for the norm on the Hilbert space and $\langle\cdot,\cdot\rangle$ for its inner product. For a map $f:A\to\mathbb R$ defined on a subset $A$ of a normed
space, we denote by
\[
\operatorname{Lip}(f)
\eqdef
\sup_{\substack{x,y\in A\\x\neq y}}
\frac{|f(x)-f(y)|}{\|x-y\|}
\in[0,\infty]
\]
its optimal Lipschitz constant. We set $\operatorname{Lip}(f)=0$ when $A$ has at most one point. Moreover, for $L \geq0$, we say that $f$ is $L$-Lipschitz whenever $\operatorname{Lip}(f)\leq L$.

Throughout, we work on a Hilbert space $L^2_{\mathbb{P}}(\mathcal{F})$, where $(\Omega,\mathcal{F},\mathbb{P})$ is a fixed probability space. We denote by $\mathbf{1}^\perp$ the orthogonal complement of the degenerate random variable 1 in $L^2_\P(\mathcal{F})$ and denote by $P_{\mathbf{1}}:L^2_{\mathbb{P}}(\mathcal{F})\to \mathbf{1}^{\perp}$ the orthogonal projection onto the orthogonal complement of $\operatorname{span}_{L^2_{\mathbb{P}}(\mathcal{F})}(\mathbf{1})$. For a map $\rho \colon L^2_{\mathbb{P}}(\mathcal{F})\to (-\infty,\infty]$, we let its reduced map $\rho^\perp$ be the restriction of $\rho$ to $\mathbf 1^\perp\cap\operatorname{dom}(\rho)$.

\subsection{Risk Measures and their Reconstruction}

We first recall the classical setup of risk measures.

\begin{definition}[Risk Measures]
    A \textit{risk measure} is a  map $\rho \colon L^2_\P(\calF)\to (-\infty, + \infty]$. It might satisfy some of the following properties 
    \begin{enumerate}[label = $\roman*)$]
        
        \item normalized: $\rho(0) = 0$.
        \item monotone: $\rho(X) \le \rho(Y)$, whenever $X \ge Y$ $\P$-a.s. and $X, Y \in L^2_\P(\calF)$.
        \item cash-additive: $\rho(X + m) = \rho(X) - m$, for all $m \in \R$ and $X \in L^2_\P(\calF)$.

        \item convex: $\rho( w X +(1-w) Y) \le w \rho(X) + (1-w) \rho(Y)$, for all $w \in [0,1]$ and $X,Y \in L^2_\P(\calF)$.

        \item  lower semicontinuous: $\rho(X) \le \liminf_{n \to \infty} \rho(X_n)$,  whenever $X\in L^2_\P(\calF)$ and $ \{X_n\}_{n \in \N}\subset L^2_\P(\calF)$ satisfy $||X_n  - X||_{L^2_\P(\calF)} \to 0$.
    \end{enumerate}
\end{definition}
\vspace{0.5em}

A risk measure $\rho$ is called a \textit{convex risk measure}, if it is monotone, cash-additive, and convex. We say a risk measure is convex if it has the property of convexity without requiring it to be a convex risk measure.

For a fixed effective number of parameters $N\in \mathbb{N}_+$, we consider data $\mathcal{D}_N$ consisting of finitely many observed elicited quotes/risk evaluation, that is
\begin{equation*}
    \mathcal{D}_N
    \eqdef
    \{(X_n,y_n)\}_{n=1}^N
    \subset
    L^2_{\mathbb{P}}(\mathcal{F})\times\mathbb{R},
\end{equation*}
where $X_n$ are contingent claims and $y_n$ are an agent's quoted valuations. Given a dataset $\mathcal{D}_N$, a natural question is whether this dataset is compatible with a risk measure, that is if the agent used one risk measure to evaluate the contingent claims. 

\begin{definition}
A risk measure $\rho$ is \textit{compatible} with data $\mathcal{D}_N$, if 
\[
\rho(X_n) = y_n \qquad \text{for all}\quad n \in [N]_+\,.
\]    
\end{definition}

We show that this question admits an answer in terms of a finite compatibility condition on the data, which we call a risk field.

\begin{definition}[Risk Field]
\label{defn:reconstruction}
A dataset $\mathcal{D}_N$
is called an $L$-\textit{risk field}, $L\geq 0$, if for every $\lambda\in \Delta_N$ and each $j\in [N]_+$,
\begin{equation}
\label{eq:whitney_riskmeasure_field__specific}
    \tilde{y}_j
    \leq
    \sum_{n=1}^N\lambda_n \tilde{y}_n
    +
    L\biggl\|
        Z_j-\sum_{n=1}^N\lambda_n\,Z_n
    \biggr\|_{L^2_{\mathbb{P}}(\mathcal{F})},
\end{equation}
where
$ Z_i\eqdef P_{\mathbf{1}}(X_i)$ and $\tilde{y}_i\eqdef y_i+\mathbb{E}[X_i]$ for all $i\in[N]_+$. 

We say that $\mathcal{D}_N$ is a
\emph{risk field} if it is an $L$-risk field for some $L\geq 0$.
\end{definition}
\vspace{0.5em}

Our next result shows that the notation of risk fields characterize datasets that are compatible with a risk measure.
 {The compatibility condition defining a risk field plays the role of Whitney's extension condition for risk measures: it completely characterizes finite datasets admitting an extension to a risk measure that is cash-additive and convex.}

\begin{theorem}[Risk Fields Characterize Universal Risk Measures]
\label{thrm: risk fields characterize}
Let $L\geq 0$ and $\mathcal D_N$ be a dataset. Then the following are equivalent:
\begin{enumerate}[(i)]
\item\label{item:existence-charac-thm}
There exists a risk measure $\rho$ that is finite valued, cash-additive, convex, compatible with $\mathcal{D}_N$, and whose restriction $\rho^\perp$ is $L$-Lipschitz.
\item\label{item:field-charac-thm}
$\mathcal D_N$ is an $L$-risk field.
\end{enumerate}
Moreover, every map $\rho$ that satisfies
\ref{item:existence-charac-thm} is globally
$\sqrt{L^2+1}$-Lipschitz continuous.
\end{theorem}

For a given dataset $\mathcal{D}_N$, we aim at finding Lipschitz continuous risk measures, parametrised by $\mathcal{D}_N$, that are compatible with the data. Specifically, we consider maps $   \varrho:L^2_{\mathbb{P}}(\mathcal{F})\times\mathbb{R}\times [0, +\infty)\times L^2_{\mathbb{P}}(\mathcal{F})\to (- \infty,  + \infty]$
sending a (data) configuration $(\mathcal{D}_N, c)$, where $c$ corresponds to the Lipschitz constant, and a random variable $Y\in L^2_{\mathbb{P}}(\mathcal{F})$ to $\rho_{\mathcal{D}_N, c}(Y)\eqdef \varrho(\mathcal{D}_N, c,Y)$.

We propose the following construction of a parametrised risk measure, which we call universal risk measure.

\begin{definition}[Universal Risk Measure]\label{def:universal_rm_construction}
For a dataset $\mathcal{D}_N$ and $c \ge 0$, the universal risk measure $\rho_{\mathcal{D}_N, c}:L^2_{\mathbb{P}}(\mathcal{F})\to (-\infty, + \infty]$ is given by
\begin{equation}
\label{eq:reconstruction_formula}
\begin{aligned}
    \rho_{\mathcal{D}_N, c}(X)
& \eqdef
    \phi(P_{\mathbf{1}}(X))-\mathbb{E}[X]\,,
    \\
\text{where} \quad
\phi(Z)
& \eqdef 
    \min_{\lambda\in\Delta_N}
    \,
    \biggl\{
        \sum_{n=1}^N \,
            \lambda_n\,\tilde{y}_n
        +
        c\,
        \biggr\|
            Z-\sum_{n=1}^N\lambda_n\,Z_n
        \biggl\|_{L^2_{\mathbb{P}}(\mathcal{F})}
    \biggr\}
\end{aligned}
\end{equation}
and $Z_n\eqdef P_\mathbf{1}(X_n) = X_n - \mathbb{E}[X_n]$ and $ \tilde{y}_n\eqdef y_n+\mathbb{E}[X_n]$, $n \in [N]_+$.
\end{definition}

Conceptually, the construction first removes the cash direction, then reconstructs a convex Lipschitz  {measure} on the quotient space, and finally restores cash-additivity.

We first show that the proposed construction given in Definition \ref{def:universal_rm_construction}, irrespective of whether there exists a risk measure that is compatible with the data, always represents a well-defined risk measure that is convex and cash-additive. This structural guarantee is independent of the choice of dataset and constant $c$.

\begin{theorem}[Certificate]
\label{thrm:Reconstruction}
Let $L\geq 0$. Then the universal risk measure $ \rho_{\mathcal{D}_N, L}(X) $ is finite-valued, cash-additive, convex, and its global Lipschitz constant satisfies $\operatorname{Lip}(\rho_{\mathcal{D}_N, L}) \le \sqrt{L^2+1}$.
\hfill\\
\noindent
Moreover, if $\mathcal{D}_N$ is an $L$-risk field, then $\rho_{\mathcal{D}_N, L}$ is compatible with $\mathcal{D}_N$. 
\end{theorem}

Whenever the universal risk measure is constructed from data $\mathcal{D}_N$ that are an $L$-risk field, and there is no ambiguity concerning $L$, we simply write $\rho_{\mathcal{D}
_N}(Y)\eqdef \rho_{\mathcal{D}_N, L}(Y) $, $Y \in L^2_{\mathbb{P}}(\mathcal{F})$.

In many applications, risk measures are required to be monotone. We show that monotonicity can be enforced by passing to the monotone hull without affecting interpolation or Lipschitz regularity.  For a monotone reconstruction, we require that the risk field is induced by a convex risk measure. Monotonicity of the quotes is necessary for the existence of a risk measure that is consistent with a risk-field. In particular the data should satisfy $y_m\le y_n$ whenever $X_n\le X_m$, $n,m\in [N]_+$, but this is in general not sufficient for monotonicity of the universal risk measure.

 {\begin{definition}
\label{defn: tamely induced dataset}
A dataset $\mathcal{D}_N$ is \textit{induced by a convex risk measure} if there exists a convex risk measure  $\rho$ that is finite-valued, compatible with $\mathcal{D}_N$, and the restriction of $\rho^{\perp} $ is Lipschitz continuous.
\end{definition}}

The next corollary shows that the monotone hull, cf.~\cite[Definition 4.1]{filipovic2007monotone}, of our reconstruction preserves all the aforementioned properties and additionally certifies monotonicity.  

\begin{corollary}[Monotone Reconstruction]
\label{cor:monotone_hull_reconstruction}
Let $L\ge0$ and $\mathcal D_N$ be compatible with a finite-valued convex risk measure $\bar\rho$ satisfying $\operatorname{Lip}(\bar\rho|_{\mathbf{1}^\perp})\le L$. Let $\rho_{\mathcal D_N,L}$ be as in~\eqref{eq:reconstruction_formula}, then
\begin{equation}
\label{eq:monotone_hull_reconstruction}
    \rho_{\mathcal D_N, L}^{\uparrow}(X)
    \eqdef
    \inf\bigl\{\rho_{\mathcal D_N, L}(Y):\,Y\le X\bigr\},
    \qquad X\in L^2_{\mathbb{P}}(\mathcal{F}).
\end{equation}
\noindent
is a convex risk measure that is $\sqrt{L^2+1}$-Lipschitz, finite-valued, monotone, cash-additive, and compatible with $\mathcal{D}_N$.
\end{corollary}

Note that we do not require that the universal risk measures are law-invariant. Thus our framework encompasses superhedging risk measures,%
\footnote{{For a discounted, locally-bounded semimartingale $S\eqdef (S_t)_{t\in [0,1]}$ on $(\Omega,\mathcal{F}_1,\mathbb{P})$, equipped with its natural filtration $\mathbb{F}\eqdef(\mathcal{F}_t)_{t\in[0,1]}$, recall that the super-hedging price $\rho(X)$ of any $X\in L^2(\Omega,\mathcal{F}_1,\mathbb{P})$ is
\[
    \rho(X)
    \eqdef
    \inf\Big\{
        m\in\mathbb{R}
        :
        \exists\pi\in\mathcal{A}_{\mathrm{bd}},
        \quad
        X+m+\int_0^1 \pi_s\,dS_s\ge0
        \quad\mathbb{P}\text{-a.s.}
    \Big\}
,
\]
with $\inf\varnothing\eqdef+\infty$. Here, $\mathcal{A}_{\mathrm{bd}}$ denotes the class of $\mathbb{F}$-predictable, $S$-integrable strategies $\pi$ whose gains processes satisfy
$
    \sup_{t\in[0,1]}
    \,|\int_0^t \pi_s\,dS_s|
    \in L^\infty(\Omega,\mathcal{F}_1,\mathbb{P})$.  
Assume that $S$ admits an equivalent local martingale measure $\mathbb{Q}$ with $d\mathbb{Q}/d\mathbb{P}\in L^2(\mathbb{P})$, ensuring that $\rho$ never takes the value $-\infty$.}%
} %
which are convex but need not be law-invariant; see~\cite[Remark~2.6(2) and Example~2.7]{zariphopoulou2010maturity}.

\subsection{Optimal Lipschitz Reconstruction}
\label{sec:optimal-Lipschitz}
We note that our proposed universal risk measure is not a simple interpolation since it has a uniformly controlled Lipschitz constant that is independent of the number of training samples. Indeed, classical universal approximation theorems~\cite{hornik1989multilayer,cybenko1989approximation} guarantee approximation on compact sets but typically do not provide quantitative control of the Lipschitz constant of the approximating network~\cite{kratsios2022do,siegel2023optimal,liu2024relu}. Such control is essential for deducing uniform approximation from interpolation on finite samples~\cite{leshno1993multilayer}.} We emphasize that our interpolation is stronger, from the perspective of regularity, than the memorization guarantees established in, e.g.,~\cite{BartlettHarveyLiawMehrabian2019VC,cuchiero2020deep,vershynin2020memory,hong2024bridging}.  

In this section, we investigate the optimality of the reconstruction formula~\eqref{eq:reconstruction_formula} from the viewpoint of Lipschitz regularity. When the data are a risk field, we introduce the smallest admissible risk-field constant $L_*(\mathcal{D}_N)$ and show that the corresponding reconstruction has the smallest Lipschitz constant among all risk measures that are finite-valued, cash-additive, convex and compatible with $\mathcal{D}_N$.

\begin{definition}[Induced dataset]
    Let $\rho$ be a risk measure. For $N\in \mathbb{N}_+$ and distinct claims $X_1,\ldots,X_N\in\operatorname{int}(\operatorname{dom}(\rho))$,
define the \textit{dataset induced by $\rho$} via
\begin{equation}
\label{eq: dataset-order-opt}
\mathcal{R}_N\eqdef\{(X_n,y_n)\}_{n=1}^N\quad \text{where}\quad y_n\eqdef\rho(X_n).
\end{equation}
\end{definition}

For simplicity, we omit the dependence of $\mathcal{R}_N$ on $\rho$.

\begin{corollary}[Lipschitz Regularity of the Reconstruction]
\label{cor:Reconstruction__orderoptimalreg}
Let $\rho$ be a risk measure that is lower semicontinuous, proper, cash-additive, and convex.  
Let $\mathcal{R}_N$ be the induced dataset by $\rho$ and define
\begin{equation}
\label{eq: intrinsic-reduced-Lipschitz}
L_{\mathcal R}
\eqdef
\operatorname{Lip}
\left(
\rho^\perp
\big|_{\operatorname{co}\{P_{\mathbf 1}(X_1),\ldots,P_{\mathbf 1}(X_N)\}}
\right).
\end{equation}
Then $L_{\mathcal R}<\infty$, $\mathcal{R}_N$ is an
$L_{\mathcal R}$-risk field, and the associated universal risk measure
$\rho_{\mathcal{R}_N}$ is finite-valued, cash-additive, convex, compatible with $\mathcal{R}_N$, and satisfies $\operatorname{Lip}(\rho_{\mathcal{R}_N})
\leq
\sqrt{L_{\mathcal R}^2+1}
$. 
\end{corollary}

The statement for globally Lipschitz risk measures then follows immediately, since the local Lipschitz constant is automatically dominated above by the global Lipschitz constant.

\begin{corollary}[Optimal Lipschitz Constant]
\label{cor: near-optimal-reconstruction}
Let $\mathcal{D}_N$ be a risk field and define
\[
L_*(\mathcal{D}_N)
\eqdef
\inf\{L\geq0:\mathcal{D}_N\text{ is an }L\text{-risk field }\}.
\]
Then $\mathcal{D}_N$ is an $L_*(\mathcal{D}_N)$-risk field. Moreover, the
universal risk measure $\rho_{\mathcal{D}_N, L_*(\mathcal{D}_N)}$ satisfies
\[
\operatorname{Lip}
\left(
\rho_{\mathcal{D}_N, L_*(\mathcal{D}_N)}|_{\mathbf{1}^\perp}
\right)
=
L_*(\mathcal{D}_N)
\]
and
\begin{align*}
\label{eq:optimal_constant}
\operatorname{Lip}(\rho_{\mathcal{D}_N, L_*(\mathcal{D}_N)})
&=
\inf
\left\{
\operatorname{Lip}(\eta):
\begin{array}{l}
\eta:L^2_{\mathbb P}(\mathcal F)\to\mathbb R
\text{ is convex and cash-additive},\\
\eta(X_n)=y_n,\quad n\in[N]_+
\end{array}
\right\}
\\
&=
\sqrt{L_*(\mathcal{D}_N)^2+1}
\end{align*}
\end{corollary}
Thus Corollary~\ref{cor: near-optimal-reconstruction} guarantees that $\rho_{\mathcal{D}_N, L_*(\mathcal{D}_N)}$ has the smallest possible global Lipschitz
constant among all risk measures that are finite-valued, cash-additive, convex, and compatible with $\mathcal{D}_N$. Next, we provide the optimal Lipschitz constant for monotone reconstructions.

\begin{corollary}[Optimal Lipschitz Constant: Monotone Version]
\label{cor:montone_hullreconstruction}
In the setting of Corollary~\ref{cor: near-optimal-reconstruction} suppose additionally that $\mathcal D_N$ is compatible with a finite-valued convex risk measure $\bar\rho$ satisfying $\operatorname{Lip}(\bar\rho|_{\mathbf{1}^\perp})\le L_*(\mathcal D_N)$. Then $
    \operatorname{Lip}(\rho_{\mathcal D_N,L_*(\mathcal D_N)}^{\uparrow})
=
    \sqrt{1+L_*(\mathcal D_N)^2}
d$.
\end{corollary}

\subsection{Consistency with Risk Profiles}
\label{sec:linear-risk-ES}

We return to the linear preferences detailed in the introduction and express the user's evaluations in the risk convention rather than the valuation convention used there.

Let $\mathbb P'$ be a probability measure on $(\Omega,\mathcal F)$ that is absolutely continuous with respect to $\mathbb P$ and denote its Radon--Nikodym derivative by $Z^\star\eqdef d\mathbb P'/d\mathbb P\in L^2_{\mathbb P}(\mathcal F)$. Consider the linear risk profile $\varrho:L^2_{\mathbb P}(\mathcal F)\to\mathbb R$ defined by 
\begin{equation}\label{eq:-linear-risk-profile}
\varrho(X)\eqdef-\mathbb E_{\mathbb P}[Z^\star X]\,.    
\end{equation}
Assume that the robo-advisor presents the user with claims having payoffs $X_1,\ldots,X_N\in L^2_{\mathbb P}(\mathcal F)$ and records the corresponding risk evaluations $y_n\eqdef\varrho(X_n)=-\mathbb E_{\mathbb P}[Z^\star X_n]$, for $n=1,\ldots,N$, that is each quote is the negative expected payoff under the user's subjective probability measure $\mathbb P'$. The resulting dataset is $\mathcal D_N\eqdef\{(X_n,y_n)\}_{n=1}^N$.

The next result confirms that the monotone reconstruction $\rho_{\mathcal D_N}^{\uparrow}$ coincides with $\varrho$ for any convex combination of the data, up to cash injections. 
Thus it perfectly reproduces the user's risk evaluations not only on the queried claims, but also on every convex combination of their payoffs and every deterministic cash shift. 

\begin{proposition}[Consistency for Linear Risk Profiles]
\label{prop:linear_quotes_monotone_hull}
Let $X_1,\ldots,X_N\in L^2_{\mathbb P}(\mathcal F)$, and set $y_n\eqdef\varrho(X_n)$, $\mathcal D_N\eqdef\{(X_n,y_n)\}_{n=1}^N$, where $\varrho$ is given in \eqref{eq:-linear-risk-profile}. Then the monotone reconstruction $\rho_{\mathcal D_N}^{\uparrow}$ with $L\eqdef\|Z^{\star}\,-\mathbf{1}\|_{L^2_{\mathbb P}(\mathcal F)}$ satisfies
\begin{equation}
\label{eq:linear_quotes_hull_exactness}
    \rho_{\mathcal D_N}^{\uparrow}\biggl(
        \sum_{n=1}^N\,
            w_n
            \,
            X_n
    \biggr)
=
    -\sum_{n=1}^N\, w_n\, \mathbb E_{\mathbb P}[Z^{\star}\,X_n]
\end{equation}
for each $w\eqdef (w_1, \ldots, w_N)\in \Delta_N$.
Moreover, the restriction of $\rho_{\mathcal D_N}^{\uparrow}$ to $\operatorname{co}\{X_1,\ldots,X_N\}+\mathbb R\mathbf{1}$ is linear.
\end{proposition}
Intuitively, Proposition~\ref{prop:linear_quotes_monotone_hull} holds because convexity and interpolation prevent the reconstruction formula from exceeding the corresponding weighted average of the user's risk quotes (which is always the case, even for datasets induced by non-linear latent risk measures). Linearity identifies this average with the user's risk evaluation.
The converse bound follows since the ``penalty term'' $L\bigl\|Z-\sum_{n=1}^N\lambda_nZ_n\bigr\|_{L^2_{\mathbb P}(\mathcal F)}$ in~\eqref{eq:reconstruction_formula}, with the prescribed choice of $L$ (namely the slope of the affine segment determined by the linear functional $\varrho$), prevents the reconstruction from dipping below that evaluation. Moreover, the $\lambda$ in the minimization may always be chosen to match any point inside the convex hull of $\{Z_n\}_{n=1}^N$, making this term vanish and making it affine on the convex hull of the queried claims. 

Next we consider non-linear preferences characterized by the Expected Shortfall (ES) risk measure. The ES is widely used following the Basel agreements and the Fundamental Review of the Trading Book (FRTB).
For $\alpha \in (0,1)$, the ES at level $\alpha$ is the map $\operatorname{ES}_{\alpha}\colon L^2(\mathcal{F})\to (-\infty,\infty]$ given by $X \in L^2(\mathcal{F})$ as
\begin{equation}
\label{eq:ES_dual_representation}
\operatorname{ES}_{\alpha}(X)\eqdef\sup_{\substack{Z\in L^2_{\mathbb P}(\mathcal F),\ \mathbb E_{\mathbb P}[Z]=1\\0\le (1-\alpha)\,Z\le 1\;\mathbb P\text{-a.s.}}}\mathbb E_{\mathbb P}[ZX].
\end{equation}

\label{ass:ES_risk_profile}
We assume the user's risk profile is $\operatorname{ES}_{\alpha}(-X)$, $\alpha \in (0,1)$. Moreover, let $X_1,\ldots,X_N\in L^2_{\mathbb P}(\mathcal F)$ and suppose there exists
$A\in\mathcal F$ and $
c_1,\dots,c_N
\in\mathbb R$ satisfying
\begin{equation}
\label{eq:tail_control}
\mathbb P(A)=1-\alpha
,\quad
\mathbb{P}(X_n \le c_n \,|\,A)=1
,\quad\mbox{ and }\quad
\mathbb P(X_n\ge c_n\,|\,   A^\complement 
)=1
\end{equation}
for every $n=1,\ldots,N$.  Set $y_n\eqdef \operatorname{ES}_{\alpha}(-X_n)$ and $\mathcal D_N\eqdef\{(X_n,y_n)\}_{n=1}^N$.

The next example illustrates the setting of European call options on a common underlying, for which case \eqref{eq:tail_control} is satisfied.
\begin{example}[European Call Options on a Common Underlying]
\label{ex:many_strikes_same_guy_striking_out}
Let $0\le K_1\le\dots\le K_N$, and $S_1\in L^2_{\mathbb P}(\mathcal F)$. Fix $\alpha\in(0,1)$, consider the quantile value 
$
    q_\alpha
    \eqdef\inf\{s\in\mathbb R:F_{S_1}(s)\ge\alpha\}
$,
and, for simplicity, assume that there is no atom at the quantile value, that is $\mathbb P(S_1=q_\alpha)=0$. Then, the short European-calls
\[
    X_n=-(S_1-K_n)_+
\]
satisfy~\eqref{eq:tail_control} with
\[
    c_n\eqdef-(q_\alpha-K_n)_+
    \quad
    \mbox{ and }
    \quad
    A\eqdef S_1^{-1}([q_\alpha,\infty)).
\]
Indeed, $\mathbb P(A)=1-\alpha$, and, $\mathbb P$-almost surely,
$X_n\le c_n$ on $A$ and $X_n\ge c_n$ on $A^\complement$.
\end{example}

\begin{proposition}[Local Consistency for ES]
\label{prop:ES_quotes_monotone_hull}
Let $X_1,\ldots,X_N\in L^2_{\mathbb P}(\mathcal F)$ satisfy \eqref{eq:tail_control} and set $y_n\eqdef\operatorname{ES}_{\alpha}(-X_n)$, $\mathcal D_N\eqdef\{(X_n,y_n)\}_{n=1}^N$. Then, the monotone reconstruction formula $\rho_{\mathcal D_N}^{\uparrow}$ in~\eqref{eq:monotone_hull_reconstruction} with $L\ge\sqrt{\alpha/(1-\alpha)}$ satisfies
\begin{equation}
\label{eq:ES_common_tail_exactness}
\begin{aligned}
\rho_{\mathcal D_N}^{\uparrow}\left(\sum_{n=1}^N w_nX_n
+H\right)
=
\operatorname{ES}_{\alpha}\left(-\sum_{n=1}^N w_nX_n
-H\right)
\end{aligned}
\end{equation}
for every $w\in\Delta_N$,
and $H\in \{Y\in L^2_{\mathbb P}(\mathcal F):Y\ge0,\ \mathbf{1}_AY=0\ \mathbb P\text{-a.s.}\}$.
\end{proposition}

The shared tail condition in~\eqref{eq:tail_control} linearizes the problem on the stated set of payoffs by allowing the supremum in the dual form of $\operatorname{ES}_{\alpha}(-X)$ in~\eqref{eq:ES_dual_representation} to be attained by the same density $\tfrac{1}{1-\alpha}\,\mathbf{1}_A$. Then, with the prescribed choice of $L$, the argument reduces to the one in Proposition~\ref{prop:linear_quotes_monotone_hull}. The additional perturbation $H$ leaves the tail evaluations unchanged, since it vanishes on $A$ and only increases payoffs outside $A$.

The lower bound on the Lipschitz constant $L\ge \sqrt{\alpha/(1-\alpha)}$ stems from the common-tail density $Z_A\eqdef(1-\alpha)^{-1}\mathbf1_A$, satisfying $\mathbb E[Z_A]=1$ and
$
\|Z_A-\mathbf1\|_{L^2_{\mathbb P}}^2
=\frac{\alpha}{1-\alpha}
$.  
Applying the Cauchy-Schwarz inequality to two arbitrary $U,V\in\mathbf1^\perp$, gives
\[
|\mathbb E[Z_A(U-V)]|
=|\mathbb E[(Z_A-\mathbf1)(U-V)]|
\le\sqrt{\frac{\alpha}{1-\alpha}}\,
\|U-V\|_{L^2_{\mathbb P}}.
\]
Thus $L\ge\sqrt{\alpha/(1-\alpha)}$ is sufficient to control the Lipschitz constant of $X\mapsto-\mathbb E[Z_AX]$ on the common tail event $A$.

\section{Numerical Illustration}
\label{sec:numerical}

To illustrate our reconstruction procedure, consider the following one-period option-risk setup. We evaluate a dual formulation of the reconstruction formula numerically using conic optimization, which allows a randomized approach to the problem's dual formulation; see Appendix~\ref{s:ToyExperiments__ss:Computation} for details.
In our experiments, the robo-adviser asks the user to quote the risk of $N=100$ short-position European call options with varying strikes; i.e.\
\[
    X_n = -(S_1-K_n)_+
    \mbox{ and }
    y_n = \rho(X_n),
    \qquad n \in [N]_+ \,,
\]
for equally spaced strikes $0<K_1<\dots<K_N$, where the user utilizes a, to the robo-adviser unknown, risk measure $\rho$.
We consider three different models for the law of $S_1$.\\

\noindent
\textbf{Geometric Brownian Motion (GBM):}
For the GBM model, we use
\[
    \mathrm{d}S_t
    =
    0.01 S_t\,\mathrm{d}t
    +
    0.01 S_t\,\mathrm{d}W_t,
    \qquad S_0=1.
\]
\textbf{Heston:}
For the Heston models, we use initial data $(S_0,v_0)=(1,0.04)$ and 
\[
    \begin{aligned}
        \mathrm{d}S_t
        &=
        0.01 S_t\,\mathrm{d}t
        +
        \sqrt{v_t}\,S_t\,\mathrm{d}W_t^{(1)},
        \\
        \mathrm{d}v_t
        &=
        2(0.04-v_t)\,\mathrm{d}t
        +
        0.3\sqrt{v_t}\,\mathrm{d}W_t^{(2)},
        \\
        \mathrm{d}\langle W^{(1)},W^{(2)}\rangle_t
        &=
        -0.7\,\mathrm{d}t.
    \end{aligned}
\]
Both continuous models are considered over the time horizon $t\in[0,1]$.\\

\noindent
\textbf{Finite State-Space Model with One Time-Step:}
For the discrete model, we assume that $S_1$ is uniformly distributed on the grid
$\{1+\tfrac{9i}{I}\}_{i=0}^{I-1}$ of $I=10^3$ equally spaced points
in $[1,10)$; i.e.,
\begin{equation}
\label{eq:Discrete_state_Space}
    \mathbb{P}\left(
        S_1 = 1+\frac{9i}{I}
    \right)
    =
    \frac{1}{I},
    \qquad i=0,\dots,I-1.
\end{equation}
\vspace{0.5em}

We consider the following risk measures in the numerical example. Let $\mathcal{P}_{\mathbb{P}}$ denote the set of probability measures on $(\Omega,\mathcal{F})$ absolutely continuous with respect to $\mathbb{P}$, and let $\Phi$ and $\Phi^{-1}$  denote the standard normal cumulative distribution function and quantile function, respectively. Then
\begin{align*}
\rho_{r}^{\mathrm{pow}}(X) 
&\eqdef \sup_{\substack{\mathbb{Q}\in\mathcal{P}_{\mathbb{P}}\\ \mathbb{Q}(A)\le 1-(1-\mathbb{P}(A))^{r}\\ \forall A\in\mathcal{F}}}\mathbb{E}_{\mathbb{Q}}[-X]
,
 \\
\rho_{\beta}^{\mathrm{exp}}(X) &\eqdef \sup_{\substack{\mathbb{Q}\in\mathcal{P}_{\mathbb{P}}\\ \mathbb{Q}(A)\le \frac{1-e^{-\beta\mathbb{P}(A)}}{1-e^{-\beta}}\\ \forall A\in\mathcal{F}}}\mathbb{E}_{\mathbb{Q}}[-X]
,
 \\
\rho_{\lambda}^{\mathrm{Wang}}(X) &\eqdef \sup_{\substack{\mathbb{Q}\in\mathcal{P}_{\mathbb{P}}\\ \mathbb{Q}(A)\le \Phi(\Phi^{-1}(\mathbb{P}(A))+\lambda)\\ \forall A\in\mathcal{F}}}\mathbb{E}_{\mathbb{Q}}[-X].
\end{align*} 
These are the latent convex risk measures considered in the application: ES, power spectral, exponential spectral, and Wang risk measures~\cite{Wang2000}. Their respective parameters are drawn uniformly at random according to
$
\alpha \sim \operatorname{Unif}\bigl([0.5,0.9)\bigr)
$ for ES, $r \sim \operatorname{Unif}\bigl([1.1,10)\bigr)$ for the power spectrum, $\beta \sim \operatorname{Unif}\bigl([0.5,10)\bigr)$ for the exponential spectrum, and $\lambda \sim \operatorname{Unif}\bigl([0.1,2)\bigr)$ for the Wang measure.

In each individual experiment, we draw parameters of the risk measures $\alpha, r, \beta, \lambda$, to create sets of risk measures the user utilizes for evaluation. Thus, each draw for each risk measure creates a new dataset of test quotes at strikes, to which we fit the universal risk measure and then report the training and test MSE.
The test MSE is evaluated on 10 additional strikes drawn uniformly from $(K_1,K_N)$.
This is repeated 30 times for each risk-measure family, and we report the mean and one sample standard deviation of these MSEs across all random parameter draws.
The strike sets are held fixed across repetitions, and the same risk-parameter draws are used across the three models.
These results are reported in Tables~\ref{tab:risk-gbm}--\ref{tab:risk-finite} below according to the ground-truth model for the underlying $S_1$.
The code is available at \url{https://github.com/AnastasisKratsios/RiskMeasureReconstruction}.

\subsection{Performance of the Reconstruction}
\label{s:ToyExperiments__ss:Performance}
The extrapolation test uses 10 new strikes, five below $K_1$ and five above $K_N$, to assess how well the reconstruction recovers the user's risk quotes beyond the observed strike range%
\footnote{%
For the GBM and Heston models, $K_1$ and $K_N$ are respectively
the $0.1$- and $0.9$-quantiles of $S_1$ under the
corresponding model. For the discrete model,
$(K_1,K_N)=(0.1,0.9)$.
The training strikes are equally spaced in $[K_1,K_N]$,
and the ten interpolation strikes are drawn uniformly
from $(K_1,K_N)$.
For the GBM and Heston models, five extrapolation strikes are drawn uniformly from each of
$[\max\{0,K_1-d\},K_1)$ and $(K_N,K_N+d]$,
where $d=(K_N-K_1)/4$.
For the discrete model, the corresponding intervals
are $[0,0.1)$ and $(0.9,1]$.
}%
.
For each parameter draw, the {out-of-distribution} (OOD) Test Ratio is the reconstruction's extrapolation MSE divided by that of the fixed ad-hoc benchmark $\mathrm{ES}_{0.99}((S_1-K)_+)$, both evaluated against the user's risk quotes at the same extrapolation strikes.
We report the mean and one sample standard deviation of these per-draw ratios; a ratio below one indicates that the reconstruction improves on the benchmark.

\begin{table}[ht]
\centering
\begin{tabular}{lccc}
\toprule
Risk measure & Train MSE & Test MSE & OOD Test Ratio \\
\midrule
ES$_\alpha$ & $0\pm0$ & $3.425\times 10^{-10}\pm 4.130\times 10^{-12}$ & $0.382 \pm 0.165$ \\
$\rho^{\mathrm{pow}}$ & $0\pm0$ & $3.445\times 10^{-10}\pm 4.463\times 10^{-12}$ & $0.374 \pm 0.135$ \\
$\rho^{\mathrm{exp}}$ & $0\pm0$ & $3.467\times 10^{-10}\pm 4.343\times 10^{-12}$ & $0.319 \pm 0.127$ \\
$\rho^{\mathrm{Wang}}$ & $0\pm0$ & $3.491\times 10^{-10}\pm 5.626\times 10^{-12}$ & $0.460 \pm 0.519$ \\
\bottomrule
\end{tabular}
\caption{GBM: mean $\pm$ one sample standard deviation over 30 parameter draws.}
\label{tab:risk-gbm}
\end{table}

\begin{table}[ht]
\centering
\begin{tabular}{lccc}
\toprule
Risk measure & Train MSE & Test MSE & OOD Test Ratio \\
\midrule
ES$_\alpha$ & $0\pm0$ & $1.159\times 10^{-7}\pm 1.405\times 10^{-9}$ & $0.473 \pm 0.200$ \\
$\rho^{\mathrm{pow}}$ & $0\pm0$ & $1.164\times 10^{-7}\pm 1.426\times 10^{-9}$ & $0.461 \pm 0.167$ \\
$\rho^{\mathrm{exp}}$ & $0\pm0$ & $1.171\times 10^{-7}\pm 1.398\times 10^{-9}$ & $0.392 \pm 0.157$ \\
$\rho^{\mathrm{Wang}}$ & $0\pm0$ & $1.178\times 10^{-7}\pm 1.873\times 10^{-9}$ & $0.580 \pm 0.681$ \\
\bottomrule
\end{tabular}
\caption{Heston: mean $\pm$ one sample standard deviation over 30 parameter draws.}
\label{tab:risk-heston}
\end{table}

\begin{table}[ht]
    \centering
    \begin{tabular}{lccc}
        \toprule
        Risk measure & Train MSE & Test MSE & OOD Test Ratio \\
        \midrule
        ES$_\alpha$ & $0\pm 0$ & $1.512\times 10^{-30}\pm 6.469\times 10^{-31}$ & $2.668\times 10^{-30}\pm 3.650\times 10^{-30}$ \\
        $\rho^{\mathrm{pow}}$ & $0\pm 0$ & $2.611\times 10^{-30}\pm 1.295\times 10^{-30}$ & $1.813\times 10^{-30}\pm 1.561\times 10^{-30}$ \\
        $\rho^{\mathrm{exp}}$ & $0\pm 0$ & $2.143\times 10^{-30}\pm 6.887\times 10^{-31}$ & $1.199\times 10^{-30}\pm 1.311\times 10^{-30}$ \\
        $\rho^{\mathrm{Wang}}$ & $0\pm 0$ & $2.033\times 10^{-30}\pm 1.225\times 10^{-30}$ & $7.159\times 10^{-31}\pm 9.337\times 10^{-31}$ \\
        \bottomrule
    \end{tabular}
    \caption{Discrete model: $S_1$ is uniformly distributed over $10^3$ equally spaced points in $[1,10)$. }
    \label{tab:risk-finite}
\end{table}

As expected, the {training} set's MSE is zero across all reported experiments and all considered user risk measures, irrespective of the market model. This is due to the fact that the datasets are induced by risk measures, and thus the universal risk measure is compatible with the data. 

In each scenario, the {test} set MSE remains very small when new strikes are drawn {within the observed strike range}; i.e.\ for randomly drawn strikes $K$ and thus previously unobserved user-quoted pairs $(X_K,\rho(X_K))$, where $K_1\le K\le K_N$, with $K_1$ and $K_N$ being the smallest and largest strikes observed in the training set, respectively. The low test set MSE is consistent with the controlled Lipschitz constant of our reconstruction and the dense training grid, which together help control interpolation error by limiting variation between nearby payoffs.

Finally, we test the OOD Test Ratio, or extrapolation, performance of our formula outside the observed strike range. This can be understood as a shift in the strike domain, while the underlying distribution and the user's risk measure remain fixed. We report the ratio of our reconstruction's test MSE on strikes sampled from the prescribed extrapolation intervals below $K_1$ and above $K_N$ to the test set MSE of the fixed benchmark $\operatorname{ES}_{0.99}$, both evaluated against the same user quotes. For both models with non-compact support, i.e.\ GBM and Heston, the mean extrapolation MSE ratios range from $0.319$ to $0.580$, corresponding to average relative MSE reductions of approximately $42\%$--$68\%$, depending on the model and risk-measure family.

\subsection{Universal Risk Measures on Incompatible Datasets}
\label{s:ToyExperiments__ss:Ablation}

We now complete our analysis by looking at datasets that are not risk-fields, thus violating the necessary and sufficient conditions for the universal risk measure to be compatible with the dataset, see Theorem~\ref{thrm: risk fields characterize}.  
For this, we fix the distribution of $S_1$ and consider a user quoting short-position European call options with payoff $(S_1-K)_+$, utilizing the preference function
\begin{align}
\label{eq:Wild_user_preferences}
y_n
&\eqdef 
    \operatorname{ES}_{\alpha}\big( (S_1-K)_+ \big) + \gamma\,  \tfrac{1-K^2}{2}\,,
\end{align}
where $0<\alpha<1$ and $0\le \gamma \le 1$. The data are thus $\{(X_n=-(S_1-K_n)_+, y_n)\}_{n=1}^N$ for uniformly sampled $0<K_1<\dots<K_N<1$ and $S_1$ is uniformly distributed on $\{1+\tfrac{9i}{I}\}_{i=0}^{I-1}$ with $I=10^3$; both in the training and test set. 

As in the numerical simulation $(S_1 - K)_+ = S_1 - K$, the preferences can be written as 
\begin{equation}\label{eq:prefernces-simple}
    \operatorname{ES}_{\alpha}\big( (S_1-K)_+ \big) + \gamma \tfrac{1-K^2}{2}
= 
    \operatorname{ES}_{\alpha}(-X) + \frac{\gamma}{2}\big(1-(S_1 + X)^2\big)\,
\end{equation}
and thus in this setting, the preferences \eqref{eq:prefernces-simple} are convex and cash additive if and only if $\gamma=0$. 
Thus, $\gamma$ might be viewed as a degree to which the required risk-field condition is violated.\footnote{For $N\ge2$ and $0<K_1<K_2<1$, the bound $S_1\ge1$ gives $X_2=X_1+(K_2-K_1)$.
Cash-additivity of $\rho$ and $\operatorname{ES}_\alpha$ therefore implies $\rho(X_2)-\rho(X_1)=\operatorname{ES}_\alpha(X_2)-\operatorname{ES}_\alpha(X_1) = - (K_2 - K_1)$. Substituting $\rho(X_n)=y_n$ yields $\frac{\gamma}{2}(K_2^2-K_1^2)=0$, contradicting $\gamma>0$; whence, no such convex risk-measure $\rho$ can exist whenever $N\ge2$.
Note that the same argument holds provided that $S_1\ge K_2$ almost surely.
}
\begin{figure}[H]%[htbp]
    \centering
    \begin{subfigure}[t]{0.48\textwidth}
        \centering
        \includegraphics[width=\linewidth]{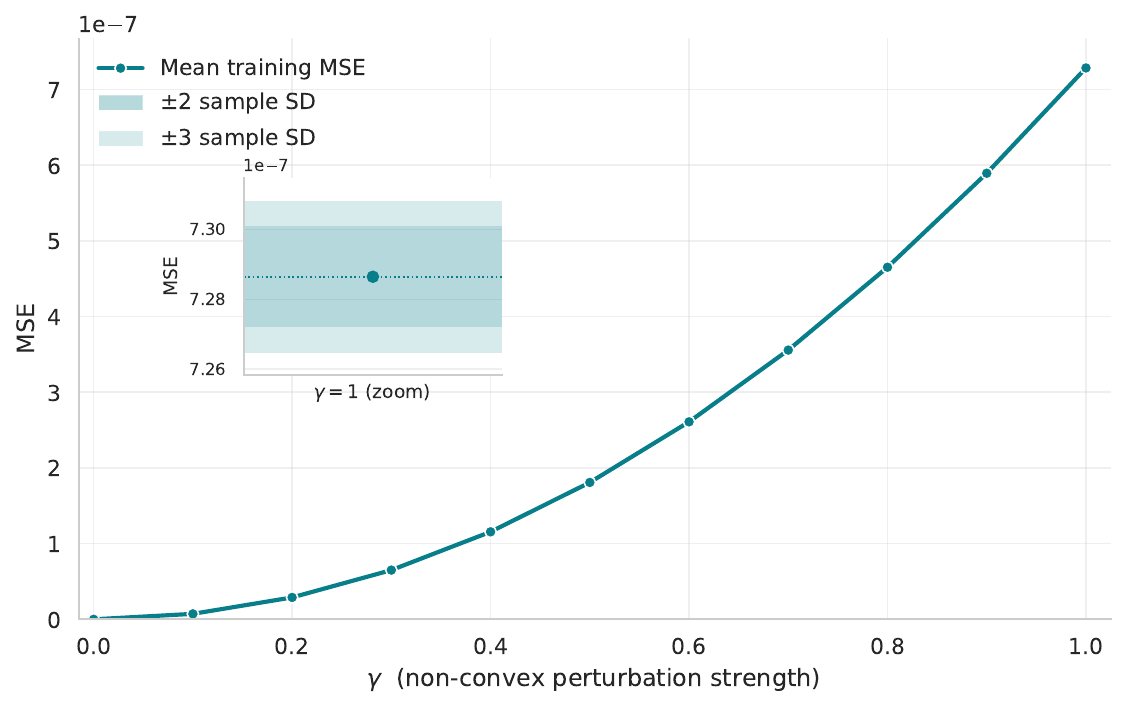}
        \caption{Training set MSE.}
    \end{subfigure}
    \hfill
    \begin{subfigure}[t]{0.48\textwidth}
        \centering
        \includegraphics[width=\linewidth]{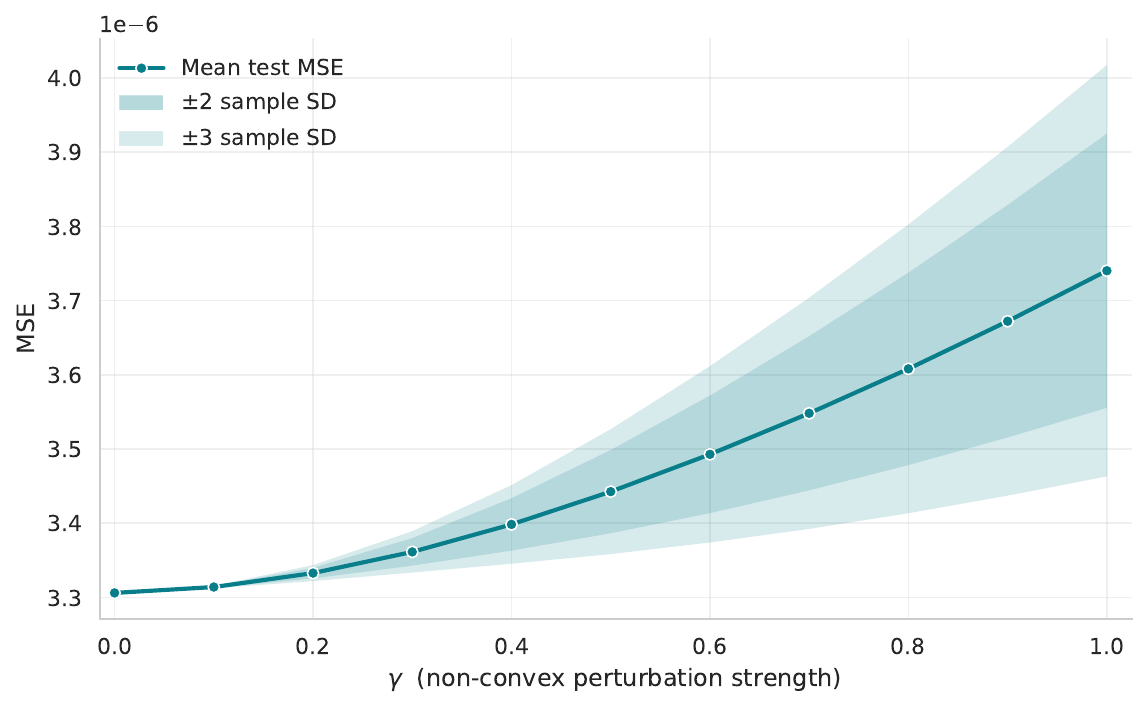}
        \caption{Test set MSE.}
    \end{subfigure}

    \caption{In-sample (training) and out-of-sample (test) performance of our reconstruction formula as the ground truth deviates from a risk-field; $\gamma=0$ yields a risk-field while $\gamma>0$ violates the risk-field condition.}
    \label{fig:concave_perturbations}
\end{figure}

As seen in Figure~\ref{fig:concave_perturbations}, the reconstruction remains accurate both in the training and test set, remaining on the order of $10^{-7}$ (resp.\ $10^{-6}$ on the test set).  As expected, the reconstruction is not exact on the training set as soon as the risk-field condition is violated ($\gamma>0$), while exact interpolation is guaranteed for the risk-field regime ($\gamma=0$).  Moreover, the test set exhibits a growing test-set variance of the reconstruction error in the non-risk-field regime, where reconstruction by the universal risk measure is not exact.

\section{Conclusion}
\label{s:Conclusion}

We study reconstruction of risk measures that are finite-valued, cash-additive, convex, and Lipschitz continuous, from finitely many evaluations of contingent claims. Viewing this as a Whitney-type extension problem with financial structure, we establish sharp compatibility conditions, called risk fields, which characterize exactly when a finite dataset admits a globally defined risk measure with the above properties. We derive an explicit reconstruction formula producing a risk measure that is finite-valued, cash-additive, convex, Lipschitz continuous and that interpolates the observed data. Moreover, choosing the smallest admissible risk-field constant yields
a Lipschitz-optimal reconstruction among all finite-valued risk measures that are cash-additive, convex, and Lipschitz continuous, and that interpolate the data. When the data are induced by a monotone risk measure, we further show that applying the monotone hull restores monotonicity without sacrificing interpolation or increasing the Lipschitz constant. 
Finally, in finite state spaces, the reconstruction admits a randomized dual implementation, and our numerical experiments demonstrate that the resulting architecture performs well in practical preference elicitation tasks.

\section{Acknowledgements}
\label{s:acks}
The authors would like to thank Damir Filipovi\'{c} and Ariel Neufeld for their helpful discussions and Rama Cont for pointing out the connection to robo advising.  
We further thank B.\ Zamanlooy for references on preference elicitation in mechanism design for algorithmic game theory and auctions.  

\section{Funding and Support}
\label{s:fund}
A.\ Kratsios acknowledges financial support from an NSERC Discovery Grant No.\ (RGPIN-2023-04482) and No.\ (DGECR-2023-00230) and by the project Bando PRIN 2022 named ``Qnt4Green - Quantitative Approaches for Green Bond Market: Risk Assessment, Agency Problems and Policy Incentives'', codice 2022JRY7EF, CUP E53D23006330006, funded by European Union – NextGenerationEU, M4c2.  
A.\ Kratsios also acknowledges that resources used in preparing this research were provided, in part, by the Province of Ontario, the Government of Canada through CIFAR, and companies sponsoring the Vector Institute\footnote{\href{https://vectorinstitute.ai/partnerships/current-partners/}{https://vectorinstitute.ai/partnerships/current-partners/}}. 
S. Pesenti acknowledges support from the NSERC Discovery Grant No.\ (RGPIN-2025-05847).
P.\ Zimmermann is supported by the Swiss National Science Foundation (SNSF) under Grant No.~239130.

\begin{appendix}
    
\section{Proofs}
\label{s:Proofs}

We first establish a lemma guaranteeing that every configuration of our formula is finite-valued, cash-additive, convex, and Lipschitz with explicit constant. Section~\ref{s:Proofs__ss:Reconstruction} contains the proof of Theorem~\ref{thrm:Reconstruction}, Section~\ref{s:Proofs__ss:Reconstruction__sss:OptimalReg} contains the proofs of Corollaries~\ref{cor:Reconstruction__orderoptimalreg} and~\ref{cor: near-optimal-reconstruction}. 
In Section~\ref{s:Proofs__ss:CorollaryMonotoneVersion}, we show that monotonicity can be restored via the monotone hull, in Section~\ref{s:Proofs__ss:Sharpness}, we prove Theorem~\ref{thrm: risk fields characterize}, and in Section~\ref{s:Proofs__ss:Applications}, we show the consistency of risk profiles discussed in Section~\ref{sec:linear-risk-ES}.

\begin{lemma}[Structural Certificate]
\label{lem:certificate}
The universal risk measure given in Definition \ref{def:universal_rm_construction} is finite-valued, cash-additive,  convex, and has a Lipschitz constant at most $\sqrt{c^2+1}$.
\end{lemma}

\begin{proof}[{Proof of Lemma~\ref{lem:certificate}}]
Fix $N\in\mathbb N_+$ and $c\geq0$, set $a_n\eqdef X_n$ and $b_n\eqdef y_n$ for $n\in[N]_+$, and write $\theta\eqdef(\{(a_n,b_n)\}_{n=1}^N,c)$.
For each $n\in[N]_+$, write
$
    Z_n^{(0)}
    \eqdef
    a_n-\mathbb{E}[a_n]\mathbf{1}
$
and
$
    \tilde b_n
    \eqdef
    b_n+\mathbb{E}[a_n]
$.
Thus $Z_n^{(0)}\in\mathbf{1}^{\perp}$ for every $n\in[N]_+$. For $\lambda\in\Delta_N$ and $Z\in\mathbf1^\perp$, set
\[
\begin{aligned}
Z^{(1)}(\lambda)(Z)&\eqdef\sum_{n=1}^N\lambda_n\tilde b_n+c\left\|Z-\sum_{n=1}^N\lambda_n Z_n^{(0)}\right\|_{L^2_{\mathbb P}(\mathcal F)},\\
Z^{(2)}(Z)&\eqdef\min_{\lambda\in\Delta_N}Z^{(1)}(\lambda)(Z)=\phi(Z),
\end{aligned}
\]
where $\phi$ is the reduced reconstruction in~\eqref{eq:reconstruction_formula}, and write $\rho_\theta\eqdef\rho_{\mathcal D_N,c}$.

 {We first observe that $Z^{(2)}$ is well defined and finite-valued.
Indeed, let $Z\in\{\mathbf{1}\}^{\perp}$ and $\lambda\in\Delta_N$.
Since $Z,\; Z_n^{(0)}\in L^2_{\mathbb P}(\mathcal F)$, we have $Z^{(1)}(\lambda)(Z)\in\mathbb R$ for every $\lambda\in\Delta_N$.
Since the map
\[
\Delta_N\ni\lambda\mapsto Z^{(1)}(\lambda)(Z)
\]
is continuous and $\Delta_N$ is compact, the minimum defining
$Z^{(2)}(Z)$ is attained. In particular, $Z^{(2)}$ is finite-valued.}

We next show that $Z^{(2)}$ is convex.  Let $Z_0,Z_1\in\mathbf{1}^{\perp}$ and let $t\in[0,1]$.  Choose $\lambda^0,\lambda^1\in\Delta_N$ attaining the minima in the definitions of $Z^{(2)}(Z_0)$ and $Z^{(2)}(Z_1)$, respectively.  Set
$
    Z_t\eqdef (1-t)Z_0+tZ_1
$
and
$
    \lambda^t\eqdef (1-t)\lambda^0+t\lambda^1
$.
Then $\lambda^t\in\Delta_N$.  Hence, by the definition of $Z^{(2)}$ and the triangle inequality,
\begin{equation}
\label{eq: convexits of Z2}
\begin{aligned}
    Z^{(2)}(Z_t)
&\le
    Z^{(1)}(\lambda^t)(Z_t)
\\
&=
    \sum_{n=1}^N\lambda^t_n\tilde b_n
    +
    c
    \left\|
        Z_t-\sum_{n=1}^N\lambda^t_n Z_n^{(0)}
    \right\|_{L^2_{\mathbb{P}}(\mathcal{F})}
\\
&=
    (1-t)\sum_{n=1}^N\lambda^0_n\tilde b_n
    +
    t\sum_{n=1}^N\lambda^1_n\tilde b_n
\\
&\qquad
    +
    c
    \left\|
        (1-t)\left(
            Z_0-\sum_{n=1}^N\lambda^0_n Z_n^{(0)}
        \right)
        +
        t\left(
            Z_1-\sum_{n=1}^N\lambda^1_n Z_n^{(0)}
        \right)
    \right\|_{L^2_{\mathbb{P}}(\mathcal{F})}
\\
&\le
    (1-t)
    \left[
        \sum_{n=1}^N\lambda^0_n\tilde b_n
        +
        c
        \left\|
            Z_0-\sum_{n=1}^N\lambda^0_n Z_n^{(0)}
        \right\|_{L^2_{\mathbb{P}}(\mathcal{F})}
    \right]
\\
&\qquad
    +
    t
    \left[
        \sum_{n=1}^N\lambda^1_n\tilde b_n
        +
        c
        \left\|
            Z_1-\sum_{n=1}^N\lambda^1_n Z_n^{(0)}
        \right\|_{L^2_{\mathbb{P}}(\mathcal{F})}
    \right]
\\
&=
    (1-t)Z^{(2)}(Z_0)+tZ^{(2)}(Z_1).
\end{aligned}
\end{equation}
Thus $Z^{(2)}$ is convex on $\mathbf{1}^{\perp}$.
We now prove that $Z^{(2)}$ is $c$-Lipschitz on $\mathbf{1}^{\perp}$.  For every $\lambda\in\Delta_N$, the map
$
    Z\mapsto Z^{(1)}(\lambda)(Z)
$
is $c$-Lipschitz, since
\[
    \left|
    \,\,
        \left\|
            Z-\sum_{n=1}^N\lambda_n Z_n^{(0)}
        \right\|_{L^2_{\mathbb{P}}(\mathcal{F})}
        -
        \left\|
            W-\sum_{n=1}^N\lambda_n Z_n^{(0)}
        \right\|_{L^2_{\mathbb{P}}(\mathcal{F})}
    \,\,
    \right|
    \le
    \|Z-W\|_{L^2_{\mathbb{P}}(\mathcal{F})}
\]
for every $Z,W\in\mathbf{1}^{\perp}$.  Therefore, for every $\lambda\in\Delta_N$, we have 
$
    Z^{(1)}(\lambda)(Z)
    \le
    Z^{(1)}(\lambda)(W)+c\|Z-W\|_{L^2_{\mathbb{P}}(\mathcal{F})}
$.
Taking the minimum over $\lambda\in\Delta_N$ gives
$
    Z^{(2)}(Z)
\le
    Z^{(2)}(W)+c\|Z-W\|_{L^2_{\mathbb{P}}(\mathcal{F})}
$.
Exchanging $Z$ and $W$ yields
\begin{equation}
\label{eq: Lipschitz of Z2}
    |Z^{(2)}(Z)-Z^{(2)}(W)|
\le
    c\|Z-W\|_{L^2_{\mathbb{P}}(\mathcal{F})}
\end{equation}
for every $Z,W\in\mathbf{1}^{\perp}$.
\hfill\\
We now pass from the reduced space $\mathbf{1}^{\perp}$ to  {$L^2_{\mathbb P}(\mathcal F)$}.  Since $P_{\mathbf{1}}$ is linear, $Z^{(2)}$ is convex, and $X\mapsto-\mathbb{E}[X]$ is  {linear}, the map
$
    \rho_\theta(X)
    =
    Z^{(2)}(P_{\mathbf{1}}(X))-\mathbb{E}[X]
$
is convex.  Since $Z^{(2)}$ is finite-valued, $\rho_\theta$ is finite-valued as well.
Next, let  {$X\in L^2_{\mathbb P}(\mathcal F)$} and $m\in\mathbb{R}$.  Since
$
    P_{\mathbf{1}}(X+m\mathbf{1})=P_{\mathbf{1}}(X)
$
and
$
    \mathbb{E}[X+m\mathbf{1}]
    =
    \mathbb{E}[X]+m
$,
we have
\[
\begin{aligned}
    \rho_\theta(X+m\mathbf{1})
&=
    Z^{(2)}(P_{\mathbf{1}}(X+m\mathbf{1}))
    -
    \mathbb{E}[X+m\mathbf{1}]
\\
&=
    Z^{(2)}(P_{\mathbf{1}}(X))
    -
    \mathbb{E}[X]
    -
    m
\\
&=
    \rho_\theta(X)-m.
\end{aligned}
\]
Thus $\rho_\theta$ is cash-additive.
\hfill\\
It remains to prove the Lipschitz estimate.  Let $X,Y\in L^2_{\mathbb P}(\mathcal F)$.  Since $Z^{(2)}$ is $c$-Lipschitz on $\mathbf{1}^{\perp}$, we find that
\[
\begin{aligned}
    |\rho_\theta(X)-\rho_\theta(Y)|
&\le
    |Z^{(2)}(P_{\mathbf{1}}(X))-Z^{(2)}(P_{\mathbf{1}}Y)|
    +
    |\mathbb{E}[X-Y]|
\\
&\le
    c\|P_{\mathbf{1}}(X-Y)\|_{L^2_{\mathbb{P}}(\mathcal{F})}
    +
    |\mathbb{E}[X-Y]|.
\end{aligned}
\]
Since $\mathbb{P}$ is a probability measure, $\|\mathbf{1}\|_{L^2_{\mathbb{P}}(\mathcal{F})}=1$, and
$
    X-Y
=
    P_{\mathbf{1}}(X-Y)+\mathbb{E}[X-Y]\mathbf{1}
$ 
is an orthogonal decomposition.  Therefore, we find that
$
    \|X-Y\|_{L^2_{\mathbb{P}}(\mathcal{F})}^2
=
    \|P_{\mathbf{1}}(X-Y)\|_{L^2_{\mathbb{P}}(\mathcal{F})}^2
+
    |\mathbb{E}[X-Y]|^2
$.
By the Cauchy-Schwarz inequality in $\mathbb{R}^2$,
\[
\begin{aligned}
    c\|P_{\mathbf{1}}(X-Y)\|_{L^2_{\mathbb{P}}(\mathcal{F})}
    +
    |\mathbb{E}[X-Y]|
&\le
    \sqrt{c^2+1}
    \left(
        \|P_{\mathbf{1}}(X-Y)\|_{L^2_{\mathbb{P}}(\mathcal{F})}^2
        +
        |\mathbb{E}[X-Y]|^2
    \right)^{1/2}
\\
&=
    \sqrt{c^2+1}\,
    \|X-Y\|_{L^2_{\mathbb{P}}(\mathcal{F})}.
\end{aligned}
\]
Hence
$
    |\rho_\theta(X)-\rho_\theta(Y)|
\le
    \sqrt{c^2+1}\,\|X-Y\|_{L^2_{\mathbb{P}}(\mathcal{F})}$ holds 
for every $X,Y\in L^2_{\mathbb P}(\mathcal F)$.
\end{proof}

\subsection{Proof of Theorem~\ref{thrm:Reconstruction}}
\label{s:Proofs__ss:Reconstruction}

Our starting point is motivated by the convex McShane-Whitney extension formula, in~\cite[Lemma 5.1]{kratsios2026structure}, which is a $0^{th}$ order version of the convex McShane-Whitney extension formula with higher-order jet data in~\cite{azagra2018explicit,azagra2017extension}.  We repeat that verbatim here.

\begin{lemma}[{\cite[Lemma 5.1]{kratsios2026structure}}]
\label{lem:convexification_finite_whitney_mcshane}
Let $H$ be a Hilbert space with norm $\|\cdot\|$, let $C\subseteq H$ be non-empty and convex, and let
$\rho:C\to\mathbb{R}$ be convex and $L$-Lipschitz.  Fix $N\in \mathbb{N}_+$ and
$\mathbb{X}_N\eqdef\{\xi_n\}_{n=1}^N\subseteq C$, and define the upper
McShane extension, cf.~\cite{cobzacs2019lipschitz}, for every $x\in H$ by
\begin{equation}
\label{eq:upper_extension}
    f_N\eqdef g_N^{\star\,\star}
\mbox{ and }
    g_N(x)
    \eqdef
    \min_{1\leq n\leq N}
    \big\{
        \rho(\xi_n)+L\|x-\xi_n\|
    \big\}.
\end{equation}
Then $f_N$ is convex, $L$-Lipschitz, and interpolates the data:
$
    f_N(\xi_n)=\rho(\xi_n)
$ for all $n=1,\dots,N$.
\hfill\\
Moreover, $f_N$ admits the semi-closed-form expression: 
\begin{equation}
\label{eq:semi_closedform}
    f_N(x)
=
    \min_{\lambda\in\Delta_N}
    \,
    \biggl\{
        \sum_{n=1}^N \lambda_n\rho(\xi_n)
        +
        L\Big\|
            x-\sum_{n=1}^N\lambda_n\xi_n
        \Big\|
    \biggr\}
.
\end{equation}
\end{lemma}
The limitation of this formula, even with computational limitations ignored, is that $f_N$ need not be a risk measure that is convex even if the ground truth risk measure $\rho$ was, since it can easily fail cash additivity.

For $\rho(x) = -x$ on $\R$, we have that $\rho$ is convex
, $1$-Lipschitz, and cash-additive in the usual payoff convention, namely $\rho(x+m)=\rho(x)-m$ for every $x,m\in\mathbb{R}$.  
However, if $N=1$ and $\xi_1=0$, then $\rho(\xi_1)=0$ and the construction gives
$
    g_1(x)=|x|
$
and hence, since $g_1$ is already convex,
$
    f_1(x)=g_1^{\star\,\star}(x)=|x|
$; where $g_1^{\star}$ (resp.\ $g_1^{\star\star}$) denotes the convex conjugate (resp., biconjugate) of $g_1$.
Thus
$
    f_1(0+1)=1\neq -1=f_1(0)-1
$,
so $f_1$ is not cash-additive.  Therefore the convexification/interpolation step preserves convexity and Lipschitz continuity, but the convex hull computed in~\eqref{eq:semi_closedform} fails to preserve the affine structure along the cash direction which is required of risk measures that are convex.   

This can be resolved by instead considering the cash-additive hull of~\cite[Remark 4.7]{filipovic2007monotone} in the cash direction.  A functional $\varrho: H\to (-\infty,\infty)$ is \textbf{$\mathfrak{c}$-cash additive}, for a fixed non-zero cash direction $\mathfrak{c}\in H$ representing one unit of cash, if for every $x\in H$ and every $a\in\mathbb{R}$ we have
\begin{equation}
\label{eq:cash_additive}
    \varrho(x+a\mathfrak{c}) = \varrho(x)-a
.
\end{equation}

We thus have the following abstract construction of a risk measure that is convex.
\begin{lemma}[Abstract Reconstruction Formula - Hilbert Space Version]
\label{lem:reconstruction_formula__convexRMabstract}
Let $H$ be a Hilbert space with norm $\|\cdot\|$, let $C\subseteq H$ be non-empty and convex, and let
$\rho:C\to\mathbb{R}$ be convex and $L$-Lipschitz.  Fix $N\in \mathbb{N}_+$ and
$\mathbb{X}_N\eqdef\{\xi_n\}_{n=1}^N\subseteq C$ and define $f_N$ by~\eqref{eq:upper_extension}--\eqref{eq:semi_closedform}.  Let $\mathfrak{c} \in H\setminus\{0\}$ be a one-unit abstract cash direction.  Assume moreover that $\rho$ admits a convex, $L$-Lipschitz, and $\mathfrak{c}$-cash-additive extension $\rho^H:H\to\mathbb{R}$.  Then, $\bar{f}_N^{\mathfrak{c}}:H\to \mathbb{R}$ defined by 
\begin{equation}
\label{eq: def of f bar}
    \bar{f}_N^{\mathfrak{c}}(x)
\eqdef
    \inf_{a\in\mathbb{R}}
    \bigl\{
        f_N(x-a\mathfrak{c})-a
    \bigl\}
\end{equation}
satisfies:
\begin{enumerate}
    \item[(i)] \textbf{Interpolation:} $\bar{f}_N^{\mathfrak{c}}(\xi_n)=\rho(\xi_n)$ for each $n=1,\dots,N$,
    \item[(ii)] \textbf{Regularity:} $\bar{f}_N^{\mathfrak{c}}$ is $L$-Lipschitz,
    \item[(iii)] \textbf{Certified risk measure that is convex:} $\bar{f}_N^{\mathfrak{c}}$ is convex and $\mathfrak{c}$-cash-additive; cf.~\eqref{eq:cash_additive}.
\end{enumerate}
\end{lemma}

 {\begin{remark}
\label{rem: lipschitz constant and cash direction}
Note that existence of $\rho^H$ in Lemma~\ref{lem:reconstruction_formula__convexRMabstract} forces $L\|\mathfrak c\|\ge 1$. Indeed, if $a\neq0$, then by
$\mathfrak c$-cash-additivity and $L$-Lipschitz continuity of $\rho^H$,
\[
|a|
=
|\rho^H(x+a\mathfrak c)-\rho^H(x)|
\le
L|a|\|\mathfrak c\|.
\]
Thus $L\|\mathfrak c\|\ge1$.
\end{remark}}

\begin{proof}[{Proof of Lemma~\ref{lem:reconstruction_formula__convexRMabstract}}]
Since $\rho^H$ is $L$-Lipschitz and
$\rho^H(\xi_n)=\rho(\xi_n)$ for every $n=1,\dots,N$, the definition of $g_N$
implies that, for every $x\in H$,
$
    \rho^H(x)\leq g_N(x)
$.
Since $f_N=g_N^{\star\,\star}$ is the largest lower-semicontinuous convex
minorant of $g_N$, and $\rho^H$ is itself a continuous convex minorant of
$g_N$, it follows that
\begin{equation}
\label{eq:rho_below_fN}
    \rho^H(x)\leq f_N(x),
    \qquad x\in H.
\end{equation}
In particular, for every $x\in H$ and every $a\in\mathbb{R}$,
\begin{equation}
\label{eq:LB1}
    f_N(x-a\mathfrak{c})-a
\geq
    \rho^H(x-a\mathfrak{c})-a
=
    \rho^H(x)
,
\end{equation}
where the last identity follows from $\mathfrak{c}$-cash-additivity of $\rho^H$.
 {Hence
$\bar{f}_N^{\mathfrak{c}}$ is finite-valued, satisfying
\begin{equation}
\label{eq: two-sided bound f bar}
    \rho^H(x)\leq \bar{f}_N^{\mathfrak{c}}(x)\leq f_N(x)
\end{equation}
for every $x\in H$, where the upper bound follows from \eqref{eq: def of f bar} with $a=0$.}

We first prove interpolation.  Fix $n\in\{1,\dots,N\}$.  
By~\eqref{eq: two-sided bound f bar}, the extension property of $\rho^H$, and \eqref{eq:semi_closedform} with $\lambda=e_n$,
$
    \rho(\xi_n)=\rho^H(\xi_n)\leq \bar{f}_N^{\mathfrak{c}}(\xi_n)\leq f_N(\xi_n)\leq \rho(\xi_n)
$; thus $\bar{f}_N^{\mathfrak{c}}(\xi_n)=\rho(\xi_n)$.
We next prove convexity.  By the preceding argument, $\bar{f}_N^{\mathfrak{c}}$
is finite-valued.  Let $x_0,x_1\in H$, let $\theta\in[0,1]$, and let
$\varepsilon>0$.  By the definition of the infimum, for each $i\in\{0,1\}$
there exists $a_i\in\mathbb{R}$ such that
$
    f_N(x_i-a_i\mathfrak{c})-a_i
    \leq
    \bar{f}_N^{\mathfrak{c}}(x_i)+\varepsilon
$.
Set $x_\theta\eqdef (1-\theta)x_0+\theta x_1$ and
$a_\theta\eqdef (1-\theta)a_0+\theta a_1$.  Then
\begin{equation}
\label{eq:ready_for_some_Yuga}
    x_\theta-a_\theta\mathfrak{c}
    =
(1-\theta)(x_0-a_0\mathfrak{c})
    +
    \theta(x_1-a_1\mathfrak{c}).
\end{equation}
Hence,~\eqref{eq:ready_for_some_Yuga} and the convexity of $f_N$ yield
\begin{equation}
\label{eq:send_doewn_to_sefr}
\begin{aligned}
    \bar{f}_N^{\mathfrak{c}}(x_\theta)
&\leq
    f_N(x_\theta-a_\theta\mathfrak{c})-a_\theta
\\
&\leq
    (1-\theta)f_N(x_0-a_0\mathfrak{c})
    +
    \theta f_N(x_1-a_1\mathfrak{c})
    -
    (1-\theta)a_0-\theta a_1
\\
&=
    (1-\theta)\bigl(f_N(x_0-a_0\mathfrak{c})-a_0\bigr)
    +
    \theta\bigl(f_N(x_1-a_1\mathfrak{c})-a_1\bigr)
\\
&\leq
    (1-\theta)\bar{f}_N^{\mathfrak{c}}(x_0)
    +
    \theta\bar{f}_N^{\mathfrak{c}}(x_1)
    +
    \varepsilon
.
\end{aligned}
\end{equation}
Since $\varepsilon>0$ was arbitrary, the desired estimate follows.

\noindent
We now prove $L$-Lipschitzness.  For every $a\in\mathbb{R}$, define
$
    F_a:H\to\mathbb{R}
$
by
$
    F_a(x)\eqdef f_N(x-a\mathfrak{c})-a
$.
Since the translation $x\mapsto x-a\mathfrak{c}$ is an isometry of $H$, and
since $f_N$ is $L$-Lipschitz, for every $x,y\in H$ we have $F_a(x)-F_a(y)
=
    f_N(x-a\mathfrak{c})-f_N(y-a\mathfrak{c})
\leq
    L\|x-y\|$;
consequently, for every $a\in\mathbb{R}$, we have
\begin{equation}
\label{eq:Fa_Lipschitz_upper}
        F_a(x)
    \leq
        F_a(y)+L\|x-y\|
.
\end{equation}
By definition,
$
    \bar{f}_N^{\mathfrak{c}}(x)=\inf_{a\in\mathbb{R}}F_a(x)
$
and
$
    \bar{f}_N^{\mathfrak{c}}(y)=\inf_{a\in\mathbb{R}}F_a(y)
$.
Taking the infimum over $a\in\mathbb{R}$ in~\eqref{eq:Fa_Lipschitz_upper}
therefore gives
\[
    \bar{f}_N^{\mathfrak{c}}(x)
    =
    \inf_{a\in\mathbb{R}}F_a(x)
    \leq
    \inf_{a\in\mathbb{R}}\bigl(F_a(y)+L\|x-y\|\bigr)
    =
    \bar{f}_N^{\mathfrak{c}}(y)+L\|x-y\|.
\]
Exchanging the roles of $x$ and $y$ yields
$
    \bar{f}_N^{\mathfrak{c}}(y)
    \leq
    \bar{f}_N^{\mathfrak{c}}(x)+L\|x-y\|
$.
Combining the two estimates gives
$
    |\bar{f}_N^{\mathfrak{c}}(x)-\bar{f}_N^{\mathfrak{c}}(y)|
    \leq
    L\|x-y\|
$.
Since $x,y\in H$ were arbitrary, $\bar{f}_N^{\mathfrak{c}}$ is
$L$-Lipschitz.
Finally, for every $b\in\mathbb{R}$,
$
    \bar{f}_N^{\mathfrak{c}}(x+b\mathfrak{c})
=
    \inf_{a\in\mathbb{R}}
    \,
    \{
        f_N(x-(a-b)\mathfrak{c})-a
    \}
$; setting $r=a-b$ gives
\[
    \bar{f}_N^{\mathfrak{c}}(x+b\mathfrak{c})
    =
    \inf_{r\in\mathbb{R}}
    \bigl\{
        f_N(x-r\mathfrak{c})-r-b
    \bigl\}
    =
    \bar{f}_N^{\mathfrak{c}}(x)-b.
\]
Thus $\bar{f}_N^{\mathfrak{c}}$ is $\mathfrak{c}$-cash-additive in the sense of~\eqref{eq:cash_additive}.
\end{proof}

We now prove our reconstruction formula.
\begin{proof}[{Proof of Theorem~\ref{thrm:Reconstruction}}]
By Lemma~\ref{lem:certificate}, $\rho_{\mathcal D_N,L}$ is finite-valued, cash-additive, convex, and $\sqrt{L^2+1}$-Lipschitz for every dataset. For the interpolation assertion, suppose that $\mathcal D_N$ is an $L$-risk field. Then condition \eqref{eq:whitney_riskmeasure_field__specific} ensures, for every $\lambda\in\Delta_N$ and every $j\in[N]_+$, that
\begin{equation}
\label{eq:risk_field_condition_shifted}
    \tilde{y}_j
    \leq
    \sum_{n=1}^N\lambda_n\tilde{y}_n
    +
    L\biggl\|
        Z_j-\sum_{n=1}^N\lambda_nZ_n
    \biggr\|_{L^2_{\mathbb{P}}(\mathcal{F})}
.
\end{equation}
Let $\phi:\mathbf{1}^{\perp}\to\mathbb{R}$ and $\rho_{\mathcal{D}_N}: L^2_{\mathbb{P}}(\mathcal{F})\to \R$ be defined as in~\eqref{eq:reconstruction_formula}. 
We first prove interpolation on the reduced space.  Fix $j\in[N]_+$.   {Taking $\lambda=e_j$ in \eqref{eq:reconstruction_formula} and using \eqref{eq:risk_field_condition_shifted} gives
\[
    \phi(Z_j)\leq \tilde{y}_j \leq
    \sum_{n=1}^N\lambda_n\tilde{y}_n
    +
    L\biggl\|
        Z_j-\sum_{n=1}^N\lambda_nZ_n
    \biggr\|_{L^2_{\mathbb{P}}(\mathcal{F})}
\]
for every $\lambda\in\Delta_N$.} Taking the minimum over $\lambda\in\Delta_N$ gives
\begin{equation}
\label{eq:phi_interpolates_shifted_values}
    \phi(Z_j)=\tilde{y}_j,
    \qquad j\in[N]_+.
\end{equation}
By \eqref{eq: convexits of Z2}--\eqref{eq: Lipschitz of Z2}, with $c=L$, $\phi$ is convex and $L$-Lipschitz. 
Thus, our next objective is to show that $\rho_{\mathcal{D}_N}$ is cash-additive.
Since $P_{\mathbf{1}}$ is linear, $\phi$ is convex, and $X\mapsto-\mathbb{E}[X]$ is linear, the function $\rho_{\mathcal{D}_N}$ is convex.  Moreover, for every $X\in L^2_{\mathbb{P}}(\mathcal{F})$ and every $a\in\mathbb{R}$,
\begin{equation}
\label{eq:lin_repeat}
    P_{\mathbf{1}}(X+a\mathbf{1})=P_{\mathbf{1}}(X)
\end{equation}
and
$
    \mathbb{E}[X+a\mathbf{1}]=\mathbb{E}[X]+a
$.
Therefore
\[
    \rho_{\mathcal{D}_N}(X+a\mathbf{1})
    =
    \phi(P_{\mathbf{1}}(X))-\mathbb{E}[X]-a
    =
    \rho_{\mathcal{D}_N}(X)-a.
\]
Thus $\rho_{\mathcal{D}_N}$ is cash-additive.
We now prove the reconstruction property.  For each $n\in[N]_+$, using~\eqref{eq:phi_interpolates_shifted_values}, we get
\[
    \rho_{\mathcal{D}_N}(X_n)
    =
    \phi(P_{\mathbf{1}}(X_n))-\mathbb{E}[X_n]
    =
    \phi(Z_n)-\mathbb{E}[X_n]
    =
    \tilde{y}_n-\mathbb{E}[X_n]
    =
    y_n.
\]
It remains to prove the Lipschitz estimate.  Let $X,Y\in L^2_{\mathbb{P}}(\mathcal{F})$. Since $\phi$ is $L$-Lipschitz,
\[
\begin{aligned}
    |\rho_{\mathcal{D}_N}(X)-\rho_{\mathcal{D}_N}(Y)|
    &\leq
    |\phi(P_{\mathbf{1}}(X))-\phi(P_{\mathbf{1}}Y)|
    +
    |\mathbb{E}[X-Y]|
    \\
    &\leq
    L\|P_{\mathbf{1}}(X-Y)\|_{L^2_{\mathbb{P}}(\mathcal{F})}
    +
    |\mathbb{E}[X-Y]|.
\end{aligned}
\]
By orthogonality, we have 
$
    X-Y=P_{\mathbf{1}}(X-Y)+\mathbb{E}[X-Y]\mathbf{1}
$; thus, 
from the proof of Lemma~\ref{lem:certificate} and the Cauchy-Schwarz inequality in $\mathbb{R}^2$, we obtain
\[
    L\|P_{\mathbf{1}}(X-Y)\|
    +
    |\mathbb{E}[X-Y]|
    \leq
    \sqrt{L^2+1}
    \bigl(
        \|P_{\mathbf{1}}(X-Y)\|_{L^2_{\mathbb{P}}(\mathcal{F})}^2
        +
        |\mathbb{E}[X-Y]|^2
    \bigr)^{1/2}.
\]
Therefore
$
    |\rho_{\mathcal{D}_N}(X)-\rho_{\mathcal{D}_N}(Y)|
    \leq
    \sqrt{L^2+1}\|X-Y\|_{L^2_{\mathbb{P}}(\mathcal{F})}
$.
Since $X,Y\in L^2_{\mathbb{P}}(\mathcal{F})$ were arbitrary, $\rho_{\mathcal{D}_N}$ is
$\sqrt{L^2+1}$-Lipschitz on $L^2_{\mathbb{P}}(\mathcal{F})$.    
\end{proof}

\subsection{Proof of Corollaries~\ref{cor:Reconstruction__orderoptimalreg} and~\ref{cor: near-optimal-reconstruction}}
\label{s:Proofs__ss:Reconstruction__sss:OptimalReg}
We now, relatively directly, obtain the sharpness of the regularity of the extension.

\begin{proof}[{Proof of Corollary~\ref{cor:Reconstruction__orderoptimalreg}}]
By cash-additivity, the domain of $\rho$ is invariant under translations in the cash direction. Indeed, $\rho(X+a\mathbf{1})=\rho(X)-a$ implies that:
\[
    X\in\operatorname{dom}(\rho)\quad\text{if and only if}\quad X+a\mathbf{1}\in\operatorname{dom}(\rho).
\]
Since translation by $a\mathbf 1$ is a homeomorphism of
$L^2_{\mathbb P}(\mathcal F)$, it follows that
\[
\operatorname{int}(\operatorname{dom}(\rho))+a\mathbf 1
=
\operatorname{int}(\operatorname{dom}(\rho)).
\]
Hence, from $X_n\in\operatorname{int}(\operatorname{dom}\rho)$, we obtain
\[
Z_n
=
P_{\mathbf 1}X_n
=
X_n-\mathbb E[X_n]\mathbf 1
\in
\operatorname{int}(\operatorname{dom}(\rho))
\cap\mathbf 1^\perp.
\]
Moreover, since $\rho$ is convex, its domain is convex,
and therefore $\operatorname{int}(\operatorname{dom}(\rho))$ is convex. Thus
\[
\operatorname{co}\{Z_1,\ldots,Z_N\}
\subseteq
\operatorname{int}(\operatorname{dom}(\rho))
\cap\mathbf 1^\perp.
\]
Since $\rho$ is proper, lower semicontinuous, and convex,
\cite[Corollary~8.30 (ii)]{AnnieLovesConvexAnalysisBook2020}
implies that
\[
\operatorname{cont}(\rho)
=
\operatorname{int}(\operatorname{dom}(\rho)).
\]
Hence, by \cite[Theorem~8.29]{AnnieLovesConvexAnalysisBook2020},
$\rho$ is locally Lipschitz on
$\operatorname{int}(\operatorname{dom}(\rho))$. Since 
\[
    \operatorname{co}\{Z_1,\ldots,Z_N\}
\subseteq
\operatorname{int}(\operatorname{dom}(\rho))
\]
is compact, local Lipschitz continuity on $\operatorname{int}(\operatorname{dom}(\rho))$ implies that there exist finitely many open balls covering $\operatorname{co}\{Z_1,\ldots,Z_N\}$
on each of which $\rho$ is Lipschitz. By the Lebesgue number lemma and a subdivision of the segment joining any two points of the convex set $\operatorname{co}\{Z_1,\ldots,Z_N\}$, these local
Lipschitz estimates combine to yield a global Lipschitz estimate. Hence
\begin{equation}
\label{eq: finite-lipschitz-order}
L_{\mathcal R}
=
\operatorname{Lip}
\left(
\rho^\perp
\big|_{\operatorname{co}\{Z_1,\ldots,Z_N\}}
\right)
<\infty.
\end{equation}

For each $n\in[N]_+$, cash-additivity and \eqref{eq: dataset-order-opt} give
\begin{equation}
\label{eq: interpolation-reduced-risk}
\rho^\perp(Z_n)
=
\rho(X_n-\mathbb E[X_n]\mathbf 1)
=
\rho(X_n)+\mathbb E[X_n]
=
\widetilde y_n,
\end{equation}
where we set
$    \widetilde y_n\eqdef y_n+\mathbb{E}[X_n]$ for $n\in [N]_+$.
Fix $j\in[N]_+$ and $\lambda\in\Delta_N$, and write
$Z_\lambda
\eqdef
\sum_{n=1}^N\lambda_n Z_n$.
By \eqref{eq: interpolation-reduced-risk} and \eqref{eq: finite-lipschitz-order},
\[
\widetilde y_j
=
\rho^\perp(Z_j)
\leq
\rho^\perp(Z_\lambda)
+
L_{\mathcal R}\|Z_j-Z_\lambda\|_{L^2_{\mathbb P}(\mathcal F)}.
\]
By convexity of $\rho$ and \eqref{eq: interpolation-reduced-risk},
\[
\rho^\perp(Z_\lambda)
\leq
\sum_{n=1}^N\lambda_n\rho^\perp(Z_n)
=
\sum_{n=1}^N\lambda_n\widetilde y_n.
\]
Consequently,
\[
\widetilde y_j
\leq
\sum_{n=1}^N\lambda_n\widetilde y_n
+
L_{\mathcal R}
\left\|
Z_j-\sum_{n=1}^N\lambda_nZ_n
\right\|_{L^2_{\mathbb P}(\mathcal F)}.
\]
Thus $\mathcal{R}_N$ is an $L_{\mathcal R}$-risk field and applying Theorem~\ref{thrm:Reconstruction} concludes the proof.
\end{proof}

Next we establish the intrinsic optimality of the reconstruction
formula with respect to global Lipschitz regularity.

\begin{proof}[{Proof of Corollary~\ref{cor: near-optimal-reconstruction}}]
For simplicity we write $\mathcal R = \mathcal{D}_N$. Since $\mathcal R$ is a risk field, the set over which the infimum is
taken in $L_*(\mathcal R)$ is non-empty, and hence $L_*(\mathcal R)<\infty$.

We first show that $\mathcal R$ is itself an
$L_*(\mathcal R)$-risk field. Let $(L_k)_{k\in\mathbb N}$ be a minimizing sequence, that is, $\mathcal R$ is an $L_k$-risk field for every $k\in\mathbb N$ and
\[
L_k\downarrow L_*(\mathcal R)\quad\text{as }k\to\infty.
\]
Without loss of generality we may assume that $(L_k)_{k\in\N}$ is monotonically decreasing. Consequently, for every $\lambda\in\Delta_N$ and every
$j\in[N]_+$,
\[
\widetilde y_j
\leq
\sum_{n=1}^N\lambda_n\widetilde y_n
+
L_k
\left\|
Z_j-\sum_{n=1}^N\lambda_nZ_n
\right\|_{L^2_{\mathbb P}(\mathcal F)}.
\]
Letting $k\to\infty$ yields
\[
\widetilde y_j
\leq
\sum_{n=1}^N\lambda_n\widetilde y_n
+
L_*(\mathcal R)
\left\|
Z_j-\sum_{n=1}^N\lambda_nZ_n
\right\|_{L^2_{\mathbb P}(\mathcal F)}.
\]
Thus $\mathcal R$ is an $L_*(\mathcal R)$-risk field.

Set $\rho_{\mathcal R}\eqdef\rho_{\mathcal D_N,L_*(\mathcal R)}$. Theorem~\ref{thrm:Reconstruction} and \eqref{eq: Lipschitz of Z2}, with $c=L_*(\mathcal R)$, imply that $\rho_{\mathcal R}$ is finite-valued, cash-additive, convex, compatible with the data, and satisfies
\begin{equation}
\label{eq: near-opt-upper}
\operatorname{Lip}(\rho_{\mathcal R}|_{\mathbf{1}^\perp})= L_*(\mathcal R)\quad\text{and}\quad\operatorname{Lip}(\rho_{\mathcal R})
\leq
\sqrt{L_*(\mathcal R)^2+1}.
\end{equation}
Indeed, these results give $\operatorname{Lip}(\rho_{\mathcal R}|_{\mathbf{1}^\perp})\leq L_*(\mathcal R)$, and equality follows by the minimality property of $L_*(\mathcal R)$. To this end, set
\[
C\eqdef
\operatorname{Lip}
\left(
\rho_{\mathcal R}|_{\mathbf{1}^\perp}
\right).
\]
Since $\rho_{\mathcal R}$ is a cash-additive convex interpolant whose
restriction to $\mathbf{1}^\perp$ is $C$-Lipschitz,
Theorem~\ref{thrm: risk fields characterize} implies that
$\mathcal R$ is a $C$-risk field. Hence, by the definition of $L_*(\mathcal R)$, $L_*(\mathcal R)\leq C$.
Consequently,
\[
\operatorname{Lip}
\left(
\rho_{\mathcal R}|_{\mathbf{1}^\perp}
\right)
=
L_*(\mathcal R).
\]

Finally, we compare $\operatorname{Lip}(\rho_{\mathcal{R}})$ with the optimal global Lipschitz constant 
\[
    K_*(\mathcal R)\eqdef \inf
\left\{
\operatorname{Lip}(\eta):
\begin{array}{l}
\eta:L^2_{\mathbb P}(\mathcal F)\to\mathbb R
\text{ is convex and cash-additive},\\
\eta(X_n)=y_n,\quad n\in[N]_+
\end{array}
\right\}.
\]
Since the risk measure $\rho_{\mathcal{R}}$ belongs to the admissible class of $K_*(\mathcal R)$, we have
\begin{equation}
\label{eq: upper bound K}
    K_*(\mathcal R)\leq \operatorname{Lip}(\rho_{\mathcal{R}})<\infty.
\end{equation}
To see the reverse inequality, choose any $\eta$ in the admissible class. If $\operatorname{Lip}(\eta)=\infty$, then \eqref{eq: upper bound K} gives $\operatorname{Lip}(\rho_{\mathcal R})\leq\operatorname{Lip}(\eta)$. It remains to treat the case when $\operatorname{Lip}(\eta)<\infty$. For this set
$K\eqdef\operatorname{Lip}(\eta|_{\mathbf1^\perp})<\infty$ and we first show that
\begin{equation}
\label{eq: exact-global-reduced-lip}
\operatorname{Lip}(\eta)
=
\sqrt{K^2+1}.
\end{equation}
Indeed, for $X,Y\in L^2_{\mathbb P}(\mathcal F)$, write
\[
X-Y
=
P_{\mathbf{1}}(X-Y)
+
\mathbb E[X-Y]\mathbf{1}.
\]
Cash-additivity implies
\[
\eta(X)
=
\eta\bigl(P_{\mathbf{1}}X\bigr)-\mathbb E[X]
\]
and similarly for $Y$. Hence
\[
\begin{aligned}
|\eta(X)-\eta(Y)|
&\leq
K
\|P_{\mathbf{1}}(X-Y)\|_{L^2_{\mathbb P}(\mathcal F)}
+
|\mathbb E[X-Y]|
\leq
\sqrt{K^2+1}\,
\|X-Y\|_{L^2_{\mathbb P}(\mathcal F)},
\end{aligned}
\]
where we used the orthogonal decomposition
\[
\|X-Y\|_{L^2_{\mathbb P}(\mathcal F)}^2
=
\|P_{\mathbf{1}}(X-Y)\|_{L^2_{\mathbb P}(\mathcal F)}^2
+
|\mathbb E[X-Y]|^2.
\]
Thus
$
\operatorname{Lip}(\eta)
\leq
\sqrt{K^2+1}$.  
If $\mathbf1^\perp=\{0\}$, then $K=0$ and $\eta(m\mathbf1)=\eta(0)-m$ for every $m\in\mathbb R$, so $\operatorname{Lip}(\eta)=1=\sqrt{K^2+1}$. Suppose now that $\mathbf1^\perp\neq\{0\}$. For the reverse inequality, let $Z,W\in\mathbf{1}^\perp$ with
$Z\neq W$, and set
\[
r\eqdef\|Z-W\|_{L^2_{\mathbb P}(\mathcal F)}>0,
\qquad
\delta\eqdef\eta(Z)-\eta(W)\in\R.
\]
For every $a\in\mathbb R$, cash-additivity gives
\[
\eta(Z+a\mathbf{1})-\eta(W)
=
\delta-a,
\]
while orthogonality yields
\[
\|Z+a\mathbf{1}-W\|_{L^2_{\mathbb P}(\mathcal F)}^2
=
r^2+a^2.
\]
Therefore
\[
\operatorname{Lip}(\eta)
\geq
\frac{|\delta-a|}{\sqrt{r^2+a^2}}
\qquad
\text{for every }a\in\mathbb R.
\]
Taking the supremum over $a\in\mathbb R$ gives
\begin{equation}
\label{eq: supremum-eta}
\operatorname{Lip}(\eta)
\geq
\sup_{a\in\mathbb R}
\frac{|\delta-a|}{\sqrt{r^2+a^2}}
=
\sqrt{1+\frac{\delta^2}{r^2}}.
\end{equation}
If $\delta=0$, then we obtain the last equality by dividing by $|a|$ and passing to the limit $|a|\to\infty$. Hence we may suppose that $\delta\neq 0$, and the change of variables $t=a/r$ and $d=\delta/r$ yield
\[
\frac{|\delta-a|}{\sqrt{r^2+a^2}}
=
\frac{|d-t|}{\sqrt{1+t^2}}
=
\frac{|\langle(d,-1),(1,t)\rangle_{\mathbb R^2}|}
     {\|(1,t)\|_{\mathbb R^2}}
\leq
\sqrt{1+d^2},
\]
where the last inequality follows from Cauchy--Schwarz. Moreover, as equality holds if and only if the vectors are linearly dependent, equality is attained for $t=-1/d$. Thus, inserting the variables $\delta$ and $r$ into \eqref{eq: supremum-eta}, we get
\[
\operatorname{Lip}(\eta)
\geq
\sqrt{
1+
\left(
\frac{|\eta(Z)-\eta(W)|}
{\|Z-W\|_{L^2_{\mathbb P}(\mathcal F)}}
\right)^2
}.
\]
Taking the supremum over all distinct $Z,W\in\mathbf{1}^\perp$ yields
\[
\operatorname{Lip}(\eta)
\geq
\sqrt{K^2+1}.
\]
Together with the reverse inequality proved above, this establishes
\eqref{eq: exact-global-reduced-lip}.
Applying \eqref{eq: exact-global-reduced-lip} to
$\eta=\rho_{\mathcal R}$ and using
\eqref{eq: near-opt-upper}, we obtain
\begin{equation}
\label{eq: exact-optimal-reconstruction-lip}
\operatorname{Lip}(\rho_{\mathcal R})
=
\sqrt{L_*(\mathcal R)^2+1}.
\end{equation}
Finally, since $\eta|_{\mathbf{1}^\perp}$ is $K$-Lipschitz, Theorem~\ref{thrm: risk fields characterize} ensures that $\mathcal R$ is a $K$-risk field. Hence, by the definition of
$L_*(\mathcal R)$,
\[
L_*(\mathcal R)\leq K.
\]
Using \eqref{eq: exact-global-reduced-lip}, we therefore obtain
\[
\operatorname{Lip}(\eta)
=
\sqrt{K^2+1}
\geq
\sqrt{L_*(\mathcal R)^2+1}
=
\operatorname{Lip}(\rho_{\mathcal R}).
\]
Taking the infimum over all admissible $\eta$ we get $\operatorname{Lip}(\rho_{\mathcal R})
\leq K_*(\mathcal R)$. Thus \eqref{eq: upper bound K} yields
\[
    \operatorname{Lip}(\rho_{\mathcal R})
= K_*(\mathcal R)
\]
and the reconstruction associated with the smallest admissible
risk-field constant is Lipschitz-optimal.
\end{proof}

\begin{proof}[{Proof of Corollary~\ref{cor:montone_hullreconstruction}}]
Corollary~\ref{cor:monotone_hull_reconstruction}, applied with $L=L_*(\mathcal D_N)$, implies that $\rho_{\mathcal D_N,L_*(\mathcal D_N)}^{\uparrow}$ is a finite-valued convex risk measure compatible with $\mathcal D_N$. Consequently, Corollary~\ref{cor: near-optimal-reconstruction} and Lemma~\ref{lem:monotone_hull_preserves_structure} give
\begin{align*}
\sqrt{1+L_*(\mathcal D_N)^2}
\le
\operatorname{Lip}\!\left(\rho_{\mathcal D_N,L_*(\mathcal D_N)}^{\uparrow}\right)
\le\operatorname{Lip}\!\left(\rho_{\mathcal D_N,L_*(\mathcal D_N)}\right)
=\sqrt{1+L_*(\mathcal D_N)^2}\,,
\end{align*}
which concludes the proof.
\end{proof}

\subsection{{Proof of Corollary~\ref{cor:monotone_hull_reconstruction}}}
\label{s:Proofs__ss:CorollaryMonotoneVersion}

\begin{lemma}[{Conservation laws of monotone hull}]
\label{lem:monotone_hull_preserves_structure}
Let $\varrho:L^2_{\mathbb{P}}(\mathcal{F})\to\mathbb{R}$ be convex, cash-additive\footnote{i.e.,
$
    \varrho(X+a\mathbf{1})=\varrho(X)-a
$ for each $X\in L^2_{\mathbb{P}}(\mathcal{F}),\ a\in\mathbb{R}$.}, and $L$-Lipschitz. % 
Let $L^2_{\mathbb P,+}(\mathcal F)\eqdef\{D\in L^2_{\mathbb P}(\mathcal F):D\geq0\ \mathbb P\text{-a.s.}\}$. Define the monotone hull of $\varrho$ by
\begin{equation}
\label{eq:monotone_hull}
    \varrho^\uparrow(X)
\eqdef
    \inf\bigl\{
        \varrho(Y):\,Y\le X
    \bigr\}
=
    \inf_{D\in L^2_{\mathbb P,+}(\mathcal F)}
        \varrho(X-D).
\end{equation}
Assume that $\varrho^\uparrow$ is finite-valued.  Then $\varrho^\uparrow$ is convex, cash-additive, monotone, and $L$-Lipschitz.  Moreover, $\varrho^\uparrow\le\varrho$, and $\varrho^\uparrow$ is the greatest monotone functional dominated by $\varrho$.
\end{lemma}

\begin{proof}[{Proof of Lemma~\ref{lem:monotone_hull_preserves_structure}}]
The two expressions in~\eqref{eq:monotone_hull} agree because $Y\le X$ if and only if $D\vcentcolon = X-Y\in L^2_{\mathbb P,+}(\mathcal F)$.  Taking $Y=X$ in \eqref{eq:monotone_hull}, yields
\begin{equation}
\label{eq:monotone_hull_below_varrho}
    \varrho^\uparrow(X)\le \varrho(X),
    \qquad X\in L^2_{\mathbb{P}}(\mathcal{F}).
\end{equation}

We first prove monotonicity.  Throughout, monotonicity of a risk measure $\rho$ is meant in the sense  {that $X\le X'$ implies that $\rho(X')\leq \rho(X)$, that is, if a portfolio $X$ results in a lower future payout (or higher loss) than another portfolio $X'$, then its risk $\rho(X)$ cannot be smaller than the risk $\rho(X')$.} If $X\le X'$, then  {the set inclusion 
\[
    \{Y\in L^2_{\mathbb{P}}(\mathcal{F}):\,Y\le X\}
\subseteq
    \{Y\in L^2_{\mathbb{P}}(\mathcal{F}):\,Y\le X'\}
\]
and \eqref{eq:monotone_hull}
imply that}
\[
    \varrho^\uparrow(X')=\inf_{Y\le X'}\varrho(Y)
\leq
\inf_{Y\le X}\varrho(Y)
        =
    \varrho^\uparrow(X).
\]
Thus $\varrho^\uparrow$ is monotone.

We next verify  {conservation of convexity}.  Let $X_0,X_1\in L^2_{\mathbb{P}}(\mathcal{F})$, let $\theta\in[0,1]$, and let $\varepsilon>0$.  Since $\varrho^\uparrow$ is finite-valued, for each $i\in\{0,1\}$ there exists $Y_i\le X_i$ such that
$
    \varrho(Y_i)
\le
    \varrho^\uparrow(X_i)+\varepsilon
$.   Set
$
    X_\theta\eqdef (1-\theta)X_0+\theta X_1
$ and $Y_\theta\eqdef (1-\theta)Y_0+\theta Y_1$.  Then, $Y_\theta\le X_\theta$ and so by \eqref{eq:monotone_hull} and the convexity of $\varrho$, we find that
\begin{equation}
\label{eq:anotherlimit_oksoooo}
\begin{aligned}
    \varrho^\uparrow(X_\theta)
    \le
    \varrho(Y_\theta)  
    \le
    (1-\theta)\varrho(Y_0)+\theta\varrho(Y_1)  
    \le
    (1-\theta)\varrho^\uparrow(X_0)+\theta\varrho^\uparrow(X_1)+\varepsilon .
\end{aligned}
\end{equation}
Letting $\varepsilon\downarrow0$ across~\eqref{eq:anotherlimit_oksoooo} establishes convexity.

Next, we show  {conservation of cash-additivity}.  Fix $X\in L^2_{\mathbb{P}}(\mathcal{F})$ and $a\in\mathbb{R}$.  The change of variables  {$W\eqdef Y-a\mathbf{1}$ and the cash-additivity of $\varrho$ imply that}
\[
\begin{aligned}
    \varrho^\uparrow(X+a\mathbf{1})
    &=
    \inf\bigl\{
        \varrho(Y):\,Y\le X+a\mathbf{1}
    \bigr\}  \\
    &=
    \inf\bigl\{
        \varrho(W+a\mathbf{1}):\,W\le X
    \bigr\}  \\
    &=
    \inf\bigl\{
        \varrho(W)-a:\,W\le X
    \bigr\}  \\
    &=
    \varrho^\uparrow(X)-a.
\end{aligned}
\]
We now check conservation of Lipschitz constant.  Indeed, for every $D\in L^2_{\mathbb P,+}(\mathcal F)$ and every $X,X'\in L^2_{\mathbb{P}}(\mathcal{F})$, the $L$-Lipschitz continuity of $\varrho$ gives
\[
    \varrho(X-D)
\le
    \varrho(X'-D)+L\|X-X'\|_{L^2_{\mathbb{P}}(\mathcal{F})}.
\] 
Taking the infimum over  {the positive cone $D\in L^2_{\mathbb P,+}(\mathcal F)$ yields
\[
    \varrho^\uparrow(X)
    \le
    \varrho^\uparrow(X')
    +
    L\|X-X'\|_{L^2_{\mathbb{P}}(\mathcal{F})}.
\]
Exchanging the roles of $X$ and $X'$ yields the reverse inequality.
Therefore
\[
|\varrho^\uparrow(X)-\varrho^\uparrow(X')|
\le
L\|X-X'\|_{L^2_{\mathbb P}(\mathcal F)},
\] 
showing the $L$-Lipschitz continuity of $\varrho^\uparrow$.}

It remains to prove the maximality assertion. Let $\psi:L^2_{\mathbb{P}}(\mathcal{F})\to\mathbb{R}\cup\{-\infty\}$ be any monotone functional satisfying $\psi\le\varrho$.  Fix $X\in L^2_{\mathbb{P}}(\mathcal{F})$.  For every $Y\le X$, monotonicity of $\psi$ gives $
    \psi(X)\le \psi(Y)
$.  Since $\psi\le\varrho$, we also have that $
    \psi(Y)\le\varrho(Y)
$.  Consequently,
\begin{equation}
\label{eq:muumuuucowsounds}
\psi(X)\le\varrho(Y)
\end{equation}
for every $Y\le X$.  Taking infima across~\eqref{eq:muumuuucowsounds} over all such $Y$ gives
$
    \psi(X)\le \varrho^\uparrow(X)
$; and since $X$ was arbitrary, $\psi\le\varrho^\uparrow$.  Therefore $\varrho^\uparrow$ is the greatest monotone functional dominated by $\varrho$.
\end{proof}
We now prove the monotone version of our main result.
\begin{proof}[{Proof of Corollary~\ref{cor:monotone_hull_reconstruction}}]
By the hypotheses on $\bar\rho$ and Theorem~\ref{thrm: risk fields characterize}, $\mathcal D_N$ is an $L$-risk field. Hence, the reconstruction formula for $\rho_{\mathcal D_N}=\rho_{\mathcal D_N,L}$ is well-defined by Theorem~\ref{thrm:Reconstruction}.
Moreover, $\bar\rho$ is finite-valued, monotone, cash-additive, convex, and compatible with data $\mathcal{D}_N$ and the restriction of $\bar \rho $ to $\mathbf{1}^{\perp}$ is $L$-Lipschitz. Define
$
    \bar\phi(Z)\eqdef \bar\rho(Z)
$
for $Z\in\mathbf{1}^{\perp}$.  
For each $n$, set $Z_n\eqdef P_{\mathbf{1}}(X_n)$ and $\tilde{y}_n\eqdef y_n+\mathbb E[X_n]$.  Since $
    X_n=Z_n+\mathbb E[X_n]\mathbf{1}
$, compatibility with $\mathcal{D}_N$ and cash-additivity of $\bar\rho$ yields
\begin{equation}
\label{eq:adding_somecash_here_and_there}
    \bar\phi(Z_n)
=
    \bar\rho(Z_n)
=
    \bar\rho(X_n)+\mathbb E[X_n]
=
    \tilde{y}_n
.
\end{equation}
Let $\phi$ denote the reduced reconstruction in~\eqref{eq:reconstruction_formula}.  Fix $Z\in\mathbf{1}^{\perp}$ and $\lambda\in\Delta_N$.  By $L$-Lipschitzness, convexity of $\bar\phi$, and~\eqref{eq:adding_somecash_here_and_there}, we find that
\begin{equation}
\label{eq:final_stepstowardswinning}
\begin{aligned}
    \bar\phi(Z)
&\le
    \bar\phi\biggl(\sum_{n=1}^N\lambda_n Z_n\biggr)
    +
    L\biggl\|Z-\sum_{n=1}^N\lambda_n Z_n\biggr\|_{L^2_{\mathbb P}(\mathcal F)}
\\
&\le
    \sum_{n=1}^N\lambda_n\bar\phi(Z_n)
    +
    L\biggl\|Z-\sum_{n=1}^N\lambda_n Z_n\biggr\|_{L^2_{\mathbb P}(\mathcal F)}  
\\
&=
    \sum_{n=1}^N\lambda_n\tilde{y}_n
    +
    L\biggl\|Z-\sum_{n=1}^N\lambda_n Z_n\biggr\|_{L^2_{\mathbb P}(\mathcal F)}.
\end{aligned}
\end{equation}
Taking the infimum across~\eqref{eq:final_stepstowardswinning} over $\lambda\in\Delta_N$ gives $\bar\phi(Z)\le\phi(Z)$.  Hence, for every $X\in L^2_{\mathbb{P}}(\mathcal{F})$, using the decomposition $X=P_{\mathbf{1}}(X)+\mathbb{E}[X]\mathbf{1}$ and the cash-additivity of $\bar\rho$, we find that
\[
    \bar\rho(X)
=
    \bar\phi(P_{\mathbf{1}}(X))-\mathbb E[X]
\le
    \phi(P_{\mathbf{1}}(X))-\mathbb E[X]
=
    \rho_{\mathcal D_N}(X);
\]that is,
\begin{equation}
\label{eq:monotone_generator_below_reconstruction}
    \bar\rho (X)
\le
    \rho_{\mathcal D_N}(X),\quad X\in L^2_{\mathbb{P}}(\mathcal{F}).
\end{equation}
We next prove that the monotone hull $\rho_{\mathcal D_N}^{\uparrow}$ satisfies all the required properties. These properties, up to interpolation of data, will follow from Lemma~\ref{lem:monotone_hull_preserves_structure}, once we have shown that \(\rho_{\mathcal D_N}^{\uparrow}\) is finite-valued, which we do next.

Since $\rho_{\mathcal D_N}^{\uparrow}\le \rho_{\mathcal D_N}$ by definition, it remains to obtain a lower bound by $\bar\rho$.  Fix $X\in L^2_{\mathbb{P}}(\mathcal{F})$.  For every $Y\le X$, monotonicity of $\bar\rho$ gives
$
    \bar\rho(X)\le \bar\rho(Y)
$.  
By~\eqref{eq:monotone_generator_below_reconstruction}, we also have $\bar\rho(Y)\le \rho_{\mathcal D_N}(Y)$.  Hence, $
    \bar\rho(X)\le \rho_{\mathcal D_N}(Y)
$ whenever $Y\le X$.  
Taking the infimum over all $Y\le X$ yields
\begin{equation}
\label{eq:monotone_generator_below_hull}
        \bar\rho(X)
    \le 
        \rho_{\mathcal D_N}^{\uparrow}(X)
.
\end{equation}
Thus,
\begin{equation}
\label{eq: unif estimate}
    \bar\rho (X)
\le
    \rho_{\mathcal D_N}^{\uparrow} (X)
\le
    \rho_{\mathcal D_N}(X),\quad X\in L^2_{\mathbb{P}}(\mathcal{F}).
\end{equation}
Now, by Theorem~\ref{thrm:Reconstruction}, $\rho_{\mathcal D_N}$ is convex, cash-additive, and $\sqrt{L^2+1}$-Lipschitz. Since $\rho_{\mathcal D_N}^{\uparrow}$ is finite-valued, Lemma~\ref{lem:monotone_hull_preserves_structure}, applied with $\varrho=\rho_{\mathcal D_N}$, implies that $\rho_{\mathcal D_N}^{\uparrow}$ is monotone, convex, cash-additive, and it satisfies
\[
    \operatorname{Lip}(\rho_{\mathcal D_N}^{\uparrow})
\le
    \operatorname{Lip}(\rho_{\mathcal D_N})
\le
    \sqrt{L^2+1}.
\]
Finally, we show that $\rho_{\mathcal D_N}^{\uparrow}$ interpolates the data. Evaluating \eqref{eq: unif estimate} at $X_n$ and using $\bar\rho(X_n)=y_n$ together with compatibility of the risk measure, we obtain
\[
    y_n
=
    \bar\rho(X_n)
\le
    \rho_{\mathcal D_N}^{\uparrow}(X_n)
\le
    \rho_{\mathcal D_N}(X_n)
=
    y_n.
\]Thus $\rho_{\mathcal D_N}^{\uparrow}(X_n)=y_n$ for every $n\in [N]_+$.
\end{proof}

\subsection{Proof of Theorem~\ref{thrm: risk fields characterize}}
\label{s:Proofs__ss:Sharpness}
If a risk measure $\rho$ interpolating the data is known to exist, then we directly obtain our static reconstruction theorem; we simply choose $H=L^2(\Omega,\mathcal{F},\mathbb{P})$ in Lemma~\ref{lem:reconstruction_formula__convexRMabstract} over a fixed probability space $(\Omega,\mathcal{F},\mathbb{P})$. The remaining main step is to characterize when paired data $\{(x_n,y_n)\}_{n=1}^N$ could have arisen from a (cash-additive) risk measure that is convex. We find a necessary and sufficient condition akin to that in~\cite{azagra2017extension}.

\begin{lemma}[Characterization of Risk Fields]
\label{lem:char-risk-fields}
Let $H$ be a Hilbert space with norm $\|\cdot \|$,
$\mathfrak c\in H\setminus\{0\}$, $L\geq 0$, and 
$\{(x_n,y_n)\}_{n=1}^N\subseteq H\times\mathbb{R}$.
Let $P_{\mathfrak c}:H\to \mathfrak c^\perp$ denote the orthogonal projection onto $\mathfrak c^\perp$. For all $x\in H$, set
\[
a_{\mathfrak c}(x)
\eqdef
\frac{\langle x,\mathfrak c\rangle}{\|\mathfrak c\|^2},
\] 
and, for each $n\in[N]_+$, define the reduced dataset $\{(z_n,\widetilde y_n)\}_{n=1}^N$ by
\begin{equation}
\label{eq: z-var-hilbert}
z_n\eqdef P_{\mathfrak c}x_n\quad\text{and}\quad \widetilde y_n
\eqdef
y_n+a_{\mathfrak c}(x_n).
\end{equation}
Then the following are equivalent:
\begin{enumerate}[(i)]
\item\label{item: extendability}
\textbf{Extendability:}
There exists a convex and $\mathfrak c$-cash additive map $\rho:H\to\mathbb{R}$, satisfying $\rho(x_n)=y_n$ for every $n\in[N]_+$, such that the restriction $\rho|_{\mathfrak c^\perp}$ is $L$-Lipschitz.

\item\label{item: risk field conditions}
\textbf{Risk-Field Conditions:}
For every $\lambda\in\Delta_N$ and each $j\in[N]_+$,
\begin{equation}
\label{eq}
\widetilde y_j
\leq
\sum_{n=1}^N\lambda_n\widetilde y_n
+
L\biggl\|
z_j-\sum_{n=1}^N\lambda_nz_n
\biggr\|.
\end{equation}
\end{enumerate}
Moreover,
any map $\rho\colon H\to \R$ satisfying the conditions in \ref{item: extendability} fulfils
\[
\operatorname{Lip}(\rho)
\leq
\sqrt{L^2+\tfrac{1}{\|\mathfrak c\|^2}}.
\]
\end{lemma}

\begin{proof}[{Proof of Lemma~\ref{lem:char-risk-fields}}]
We first prove that \ref{item: extendability} implies
\ref{item: risk field conditions}. Assume that there exists a convex, $\mathfrak c$-cash additive map $\rho:H\to\mathbb{R}$ satisfying $\rho(x_n)=y_n$ for every $n\in[N]_+$, whose restriction to $\mathfrak c^\perp$ is $L$-Lipschitz.
Since the orthogonal projection $P_{\mathfrak c}$ can be explicitly written as $P_{\mathfrak c}=\textrm{id}-\mathrm{c}\,a_{\mathfrak c}$, every $x\in H$ can be decomposed as
\begin{equation}
\label{eq: decomposition H}
x
=
P_{\mathfrak c}x
+
a_{\mathfrak c}(x)\mathfrak c
.
\end{equation}
and the $\mathfrak c$-cash additivity of $\rho$ implies
\begin{equation}
\label{eq: decomp-rho-hilbert}
\rho(x)
=
\rho(P_{\mathfrak c}x)
-
a_{\mathfrak c}(x)
.
\end{equation}
If we denote the restriction $\rho|_{\mathfrak c^\perp}\colon \mathfrak c^\perp\to \R$ by $\phi$, then the previous formula becomes
\begin{equation}
\label{eq: hilbert-phi}
\rho(x)
=
\phi(P_{\mathfrak c}x)-a_{\mathfrak c}(x),
\qquad x\in H.
\end{equation}
In particular, for every $n\in[N]_+$, \eqref{eq: z-var-hilbert}, \eqref{eq: hilbert-phi}, and $y_n=\rho(x_n)$ yield
\begin{equation}
\label{eq: interpolation phi hilbert}
    \phi(z_n)=\phi(P_{\mathfrak c}x_n)=\rho(x_n)+a_{\mathfrak c}(x_n)=y_n+a_{\mathfrak c}(x_n)=\widetilde y_n.
\end{equation}
Thus $\phi$ interpolates the reduced data
$\{(z_n,\widetilde y_n)\}_{n=1}^N$.
\hfill\\
\noindent Note that by the properties of $\rho$, $\phi$ is convex and $L$-Lipschitz on $\mathfrak c^\perp$. We now prove property~\eqref{eq}. Fix an integer $j\in[N]_+$ and let
\begin{equation}
\label{eq: zlambda}
z_\lambda\eqdef \sum_{n=1}^N\lambda_n z_n,\quad \lambda\in\Delta_N.
\end{equation}
By $L$-Lipschitzness and the interpolation property of $\phi$,
we get
\[
\widetilde y_j
=
\phi(z_j)=(\phi(z_j)-\phi(z_\lambda))+\phi(z_\lambda)
\leq
\phi(z_\lambda)+L\|z_j-z_\lambda \|.
\]
Hence \eqref{eq: zlambda}, convexity of $\phi$, and \eqref{eq: interpolation phi hilbert} yield
\begin{equation*}
\widetilde y_j \leq \phi\Bigg(\sum_{n=1}^N\lambda_nz_n\Bigg)+L\|z_j-z_\lambda \|
\leq
\sum_{n=1}^N\lambda_n\phi(z_n)+L\|z_j-z_\lambda \|
=
\sum_{n=1}^N\lambda_n\widetilde y_n+L\|z_j-z_\lambda \|.   
\end{equation*}
This establishes~\eqref{eq}.
\hfill\\
\noindent
Conversely, assume \ref{item: risk field conditions} holds and define
$\phi:\mathfrak c^\perp\to\mathbb{R}$ by the convex McShane extension formula
\[
\phi(z)
\eqdef
\min_{\lambda\in\Delta_N}
\biggl\{
\sum_{n=1}^N\lambda_n\widetilde y_n
+
L\biggl\|
z-\sum_{n=1}^N\lambda_nz_n
\biggr\|
\biggr\}.
\]
Note that the same reasoning as in the proof of
Theorem~\ref{thrm:Reconstruction}, dealing with the special case
$H=L^2_{\mathbb{P}}(\mathcal{F})$, carries over to the present
setting, which demonstrates that $\phi$ is well-defined and interpolates the reduced data, i.e.
$\phi(z_j)=\widetilde y_j$, for every $j\in[N]_+$.
\hfill\\
\noindent
The convexity and Lipschitz property of $\phi$ follow exactly as in equations \eqref{eq: convexits of Z2}--\eqref{eq: Lipschitz of Z2} in the proof of Lemma~\ref{lem:certificate} or as in the
proof of Theorem~\ref{thrm:Reconstruction}. Indeed, let
$z_0,z_1\in\mathfrak c^\perp$ and $\theta\in[0,1]$.
For each $i\in\{0,1\}$, choose $\lambda^i\in\Delta_N$ such that
\[
\phi(z_i)
=
\sum_{n=1}^N\lambda^i_n\widetilde y_n
+
L\biggl\|
z_i-\sum_{n=1}^N\lambda^i_nz_n
\biggr\|.
\]
Set
$
z_\theta\eqdef (1-\theta)z_0+\theta z_1
$
and
$
\lambda^\theta\eqdef
(1-\theta)\lambda^0+\theta\lambda^1$.  
Since $\lambda^\theta\in\Delta_N$, we have
\begin{align*}
\phi(z_\theta)
&\leq
\sum_{n=1}^N\lambda^\theta_n\widetilde y_n
+
L\biggl\|
z_\theta-\sum_{n=1}^N\lambda^\theta_nz_n
\biggr\|
\\
&=
(1-\theta)\sum_{n=1}^N\lambda^0_n\widetilde y_n
+
\theta\sum_{n=1}^N\lambda^1_n\widetilde y_n
+
L\biggl\|
(1-\theta)
\biggl(z_0-\sum_{n=1}^N\lambda^0_nz_n\biggr)
+
\theta
\biggl(z_1-\sum_{n=1}^N\lambda^1_nz_n\biggr)
\biggr\|
\\
&\leq
(1-\theta)
\biggl[
\sum_{n=1}^N\lambda^0_n\widetilde y_n
+
L\biggl\|
z_0-\sum_{n=1}^N\lambda^0_nz_n
\biggr\|
\biggr]
+
\theta
\biggl[
\sum_{n=1}^N\lambda^1_n\widetilde y_n
+
L\biggl\|
z_1-\sum_{n=1}^N\lambda^1_nz_n
\biggr\|
\biggr]
\\
&=
(1-\theta)\phi(z_0)+\theta\phi(z_1)\,,
\end{align*}
which proves that $\phi$ is convex.

Next, we prove that $\phi$ is $L$-Lipschitz. For every
$\lambda\in\Delta_N$, set
\[
F_\lambda:\mathfrak c^\perp\to\mathbb{R},\quad F_\lambda(z)
\eqdef
\sum_{n=1}^N\lambda_n\widetilde y_n
+
L\biggl\|
z-\sum_{n=1}^N\lambda_nz_n
\biggr\|.
\]
Then, by the triangle inequality, each $F_\lambda$ is $L$-Lipschitz.
Hence, for every $z,w\in\mathfrak c^\perp$ and every
$\lambda\in\Delta_N$,
$
F_\lambda(z)\leq F_\lambda(w)+L\|z-w\|$.  
Taking the infimum over $\lambda\in\Delta_N$ gives
$
\phi(z)\leq \phi(w)+L\|z-w\|$.
Exchanging $z$ and $w$ yields
$
\phi(w)\leq \phi(z)+L\|z-w\|
$
and therefore
\[
|\phi(z)-\phi(w)|\leq L\|z-w\|.
\]
Thus $\phi$ is $L$-Lipschitz.  Finally, define $\rho:H\to\mathbb{R}$ by
\begin{equation}
\label{eq: def-rho-hilbert}
\rho(x)
\eqdef
\phi(P_{\mathfrak c}x)-a_{\mathfrak c}(x).
\end{equation}
Now, similarly to the argument circa~\eqref{eq:lin_repeat}, since $P_{\mathfrak c}$ is linear, $\phi$ is convex, and $x\mapsto-a_{\mathfrak c}(x)$ is linear, the map $\rho$ is convex.

Moreover, for every $z\in\mathfrak c^\perp$ we have $a_{\mathfrak c}(z)=0$ and $P_{\mathfrak c}z=z$, implying that $\rho(z)=\phi(z)$ for $z\in\mathfrak c^\perp$.  Consequently, by construction, the restriction $\rho|_{\mathfrak c^\perp}$ is $L$-Lipschitz.
Next, let $m\in \R$. Since
$
P_{\mathfrak c}(x+m\mathfrak c)=P_{\mathfrak c}x
$ and $
a_{\mathfrak c}(x+m\mathfrak c)
=
a_{\mathfrak c}(x)+m
$; then we have
\[
\begin{aligned}
\rho(x+m\mathfrak c)
&=
\phi(P_{\mathfrak c}(x+m\mathfrak c))
-
a_{\mathfrak c}(x+m\mathfrak c)
\\
&=
\phi(P_{\mathfrak c}x)-a_{\mathfrak c}(x)-m
\\
&=
\rho(x)-m.
\end{aligned}
\]
Thus $\rho$ is $\mathfrak c$-cash additive (see~\eqref{eq:cash_additive}).
\hfill\\
\noindent
Finally, for every $n\in[N]_+$, \eqref{eq: z-var-hilbert} and $\phi(z_j)=\widetilde y_j$ yield
\[
\rho(x_n)
=
\phi(P_{\mathfrak c}x_n)-a_{\mathfrak c}(x_n)=
\phi(z_n)-a_{\mathfrak c}(x_n)
=
\widetilde y_n-a_{\mathfrak c}(x_n)
=
y_n.
\]
This proves~\ref{item: extendability}.
\hfill\\
\noindent
It remains to establish the asserted global Lipschitz
bound. Let $x,x'\in H$. Equation \eqref{eq: decomp-rho-hilbert}, the linearity of $a_{\mathfrak c}$, the decomposition
\[
    x= P_{\mathfrak  c}x+a_{\mathfrak c}(x)\mathfrak c,
\]
the triangle inequality, and the $L$-Lipschitz continuity of $\rho|_{\mathfrak c^\perp}$ guarantee that
\[
\begin{split}
|\rho(x)-\rho(x')|&=|\rho(P_{\mathfrak c} x)-\rho(P_{\mathfrak c} x')-a_{\mathfrak c}(x)+a_{\mathfrak c}(x')|\\
&\leq
L\|P_{\mathfrak c}(x-x')\| +|a_{\mathfrak c}(x-x')|
\end{split}
\]
Moreover, since $P_{\mathfrak c}$ is an orthogonal projection, we have
\[
\|x-x'\|^2
=
\|P_{\mathfrak c}(x-x')\|^2
+
|a_{\mathfrak c}(x-x')|^2\|\mathfrak c\|^2.
\]
Note that applying the Cauchy--Schwarz inequality in
$\mathbb{R}^2$ yields
\[
\begin{aligned}
L\|P_{\mathfrak c}(x-x')\|
+
|a_{\mathfrak c}(x-x')|
&
\leq
\sqrt{L^2+\tfrac{1}{\|\mathfrak c\|^2}}
\,
\bigl(
\|P_{\mathfrak c}(x-x')\|^2
+
|a_{\mathfrak c}(x-x')|^2\|\mathfrak c\|^2
\bigr)^{1/2}
\\
&
=
\sqrt{L^2+\tfrac{1}{\|\mathfrak c\|^2}}
\,\|x-x'\|.
\end{aligned}
\]
Therefore we deduce that
$
\operatorname{Lip}(\rho)
\leq
\sqrt{L^2+\tfrac{1}{\|\mathfrak c\|^2}}$, 
as claimed.
\end{proof}

We now give the proof of Theorem~\ref{thrm: risk fields characterize}.

\begin{proof}[Proof of Theorem~\ref{thrm: risk fields characterize}]
We apply Lemma~\ref{lem:char-risk-fields} with $H\eqdef L^2_{\mathbb P}(\mathcal F)$ and $\mathfrak c\eqdef \mathbf{1}$.  Since $\mathbb P$ is a probability measure,
$
\|\mathbf{1}\|_{L^2_{\mathbb P}(\mathcal F)}^2
=
\mathbb E[\mathbf{1}]
=
1$ and 
$a_{\mathbf{1}}(X)
=
\tfrac{\langle X,\mathbf{1}\rangle_{L^2_{\mathbb P}}}
{\|\mathbf{1}\|_{L^2_{\mathbb P}}^2}
=
\mathbb E[X]
$; 
for every $X\in L^2_{\mathbb P}(\mathcal F)$. The orthogonal projection $P_{\mathfrak c}$ appearing in
Lemma~\ref{lem:char-risk-fields} is therefore precisely the projection
\[
P_{\mathbf{1}} X=X-\mathbb E[X]\mathbf{1}
\]
onto $\mathbf{1}^{\perp}$.
Consequently, the reduced data from
\eqref{eq: z-var-hilbert} are
\[
Z_n=P_{\mathbf{1}}(X_n)
\qquad\text{and}\qquad
\widetilde y_n
=
y_n+\mathbb E[X_n],
\]
which are exactly the quantities appearing in Definition~\ref{defn:reconstruction}.
Thus the risk-field condition in Lemma~\ref{lem:char-risk-fields} becomes
\[
\widetilde y_j
\leq
\sum_{n=1}^N\lambda_n\widetilde y_n
+
L
\left\|
Z_j-\sum_{n=1}^N\lambda_nZ_n
\right\|_{L^2_{\mathbb P}(\mathcal F)}
\]
for every $\lambda\in\Delta_N$ and every $j\in[N]_+$, which is precisely the condition that $\mathcal D_N$ is an $L$-risk field. Hence Lemma~\ref{lem:char-risk-fields} yields the equivalence between \ref{item:existence-charac-thm} and \ref{item:field-charac-thm}.
Finally, the last assertion of Lemma~\ref{lem:char-risk-fields}, together with
$\|\mathbf 1\|_{L^2_{\mathbb P}(\mathcal F)}=1$, gives
$
\operatorname{Lip}(\rho)^2
\leq
L^2+\big(\|\mathbf 1\|_{L^2_{\mathbb P}(\mathcal F)}^2\big)^{-1}
=
L^2+1
$.
\end{proof}

\subsection{Proofs of Propositions~\ref{prop:linear_quotes_monotone_hull} and~\ref{prop:ES_quotes_monotone_hull}}
\label{s:Proofs__ss:Applications}

\begin{proof}[{Proof of Proposition~\ref{prop:linear_quotes_monotone_hull}}]
Write $\mathbb E\eqdef\mathbb E_{\mathbb P}$, $\|\cdot\|\eqdef\|\cdot\|_{L^2_{\mathbb P}(\mathcal F)}$, and $\langle U,V\rangle\eqdef\mathbb E[UV]$. Put $a\eqdef\mathbf{1}-Z^{\star}\,$. Since $Z^{\star}\,\ge0$ and $\mathbb E[Z^{\star}\,]=1$, we have $\mathbb E[a]=0$ and $\|a\|=L$. For every $X\in L^2_{\mathbb P}(\mathcal F)$, Cauchy--Schwarz gives $\mathbb E[|Z^{\star}\,X|]\le\|Z^{\star}\,\|\|X\|<\infty$, and $\varrho(X)=-\mathbb E[X]+\langle a,P_{\mathbf{1}}(X)\rangle$.

As in Theorem~\ref{thrm:Reconstruction}, set $Z_n\eqdef P_{\mathbf{1}}(X_n)$ and $\tilde y_n\eqdef y_n+\mathbb E[X_n]=\langle a,Z_n\rangle$. For $\lambda\in\Delta_N$, write $Z_\lambda\eqdef\sum_{n=1}^N\lambda_nZ_n$. For every $j=1,\ldots,N$ and $\lambda\in\Delta_N$,
\[
\tilde y_j-\sum_{n=1}^N\lambda_n\tilde y_n
=\langle a,Z_j-Z_\lambda\rangle
\le\|a\|\|Z_j-Z_\lambda\|
=L\|Z_j-Z_\lambda\|.
\]
Thus $\mathcal D_N$ is an $L$-risk field. For each $X\in L^2_{\mathbb P}(\mathcal F)$, the objective defining $\rho_{\mathcal D_N}(X)$ is real-valued and continuous on the nonempty compact simplex $\Delta_N$, so its minimum exists and is finite. Substitution into the reconstruction formula of Theorem~\ref{thrm:Reconstruction} gives
\begin{align*}
\rho_{\mathcal D_N}(X)-\varrho(X)
&=\min_{\lambda\in\Delta_N}\left\{L\|P_{\mathbf{1}}(X)-Z_\lambda\|-\langle a,P_{\mathbf{1}}(X)-Z_\lambda\rangle\right\}\\
&\ge\min_{\lambda\in\Delta_N}\left\{(L-\|a\|)\|P_{\mathbf{1}}(X)-Z_\lambda\|\right\}=0.
\end{align*}
If $Y\in L^2_{\mathbb P}(\mathcal F)$ satisfies $Y\le X$ $\mathbb P$-a.s., then
\[
\rho_{\mathcal D_N}(Y)\ge\varrho(Y)=\varrho(X)+\mathbb E[Z^{\star}\,(X-Y)]\ge\varrho(X).
\]
Taking the infimum over such $Y$ in~\eqref{eq:monotone_hull_reconstruction} and using the admissible choice $Y=X$ yields
\begin{equation}
\label{eq:linear_quotes_hull_sandwich}
-\infty<\varrho(X)\le\rho_{\mathcal D_N}^{\uparrow}(X)\le\rho_{\mathcal D_N}(X)<\infty,
\qquad X\in L^2_{\mathbb P}(\mathcal F).
\end{equation}

Fix $\lambda^\star\in\Delta_N$ and $c\in\mathbb R$, and set $X\eqdef\sum_{n=1}^N\lambda_n^\star X_n+c\mathbf{1}$. Since $P_{\mathbf{1}}(X)=Z_{\lambda^\star}$ and $\mathbb E[X]=\sum_{n=1}^N\lambda_n^\star\mathbb E[X_n]+c$, evaluating the reconstruction objective at $\lambda^\star$ gives
\begin{align*}
\rho_{\mathcal D_N}(X)
&\le\sum_{n=1}^N\lambda_n^\star\tilde y_n-\mathbb E[X]\\
&=\sum_{n=1}^N\lambda_n^\star\bigl(y_n+\mathbb E[X_n]\bigr)-\sum_{n=1}^N\lambda_n^\star\mathbb E[X_n]-c\\
&=\sum_{n=1}^N\lambda_n^\star\varrho(X_n)-c
=-\mathbb E\left[Z^{\star}\,\left(\sum_{n=1}^N\lambda_n^\star X_n+c\mathbf{1}\right)\right]
=\varrho(X).
\end{align*}
Together with~\eqref{eq:linear_quotes_hull_sandwich} this proves~\eqref{eq:linear_quotes_hull_exactness}.
\end{proof}

\begin{proof}[{Proof of Proposition~\ref{prop:ES_quotes_monotone_hull}}]
Write $\mathbb E\eqdef\mathbb E_{\mathbb P}$, $\|\cdot\|\eqdef\|\cdot\|_{L^2_{\mathbb P}(\mathcal F)}$, and $\langle U,V\rangle\eqdef\mathbb E[UV]$. Let $\mathcal Q_\alpha$ denote the set of densities in~\eqref{eq:ES_dual_representation}, and put $Z_A\eqdef(1-\alpha)^{-1}\mathbf{1}_A$. Since $0\le Z_A\le(1-\alpha)^{-1}$ and $\mathbb E[Z_A]=(1-\alpha)^{-1}\mathbb P(A)=1$, we have $Z_A\in\mathcal Q_\alpha$. Every $\zeta\in\mathcal Q_\alpha$ satisfies $\|\zeta\|^2=\mathbb E[\zeta^2]\le(1-\alpha)^{-1}\mathbb E[\zeta]=(1-\alpha)^{-1}$. Consequently, $\sup_{\zeta\in\mathcal Q_\alpha}|\mathbb E[\zeta U]|\le(1-\alpha)^{-1/2}\|U\|<\infty$ for every $U\in L^2_{\mathbb P}(\mathcal F)$.
\hfill\\
\noindent
Fix $U\in L^2_{\mathbb P}(\mathcal F)$ and $d\in\mathbb R$ such that $U\le d$ on $A$ and $U\ge d$ on $A^c$, $\mathbb P$-a.s. For every $\zeta\in\mathcal Q_\alpha$,
\begin{align*}
\mathbb E[(\zeta-Z_A)U]
&=\mathbb E[(\zeta-Z_A)(U-d\mathbf{1})]\\
&=\mathbb E\big[\mathbf{1}_A\big(\zeta-\tfrac1{1-\alpha}\big)(U-d\mathbf{1})\big]+\mathbb E[\mathbf{1}_{A^c}\zeta(U-d\mathbf{1})]\ge0.
\end{align*}
The first equality uses $\mathbb E[\zeta-Z_A]=0$, and both integrands in the second line are nonnegative. Thus $-\mathbb E[\zeta U]\le-\mathbb E[Z_AU]$, with equality at $\zeta=Z_A$. By~\eqref{eq:ES_dual_representation},
\begin{equation}
\label{eq:ES_common_tail_linearization}
\operatorname{ES}_{\alpha}(-U)=-\mathbb E[Z_AU]
\end{equation}
whenever $U$ satisfies these inequalities. Since $\mathbb P(A)=1-\alpha>0$ and $\mathbb P(A^c)=\alpha>0$, the conditional-probability identities in~\eqref{eq:tail_control} give $X_n\le 
c_n
$ on $A$ and $X_n\ge 
c_n
$ on $A^c$, $\mathbb P$-a.s. Hence $y_n\eqdef\operatorname{ES}_{\alpha}(-X_n)=-\mathbb E[Z_AX_n]$ for every $n=1,\ldots,N$.
\hfill\\
\noindent
Put $a\eqdef\mathbf{1}-Z_A$. Then $\mathbb E[a]=0$ and
\[
\|a\|^2=\mathbb E[Z_A^2]-1=\tfrac{\mathbb P(A)}{(1-\alpha)^2}-1=\tfrac{\alpha}{1-\alpha}\le L^2.
\]
As in Theorem~\ref{thrm:Reconstruction}, set $Z_n\eqdef P_{\mathbf{1}}(X_n)$ and $\tilde y_n\eqdef y_n+\mathbb E[X_n]=\langle a,Z_n\rangle$. For $\lambda\in\Delta_N$, write $Z_\lambda\eqdef\sum_{n=1}^N\lambda_nZ_n$. For every $j=1,\ldots,N$ and $\lambda\in\Delta_N$,
\[
\tilde y_j-\sum_{n=1}^N\lambda_n\tilde y_n=\langle a,Z_j-Z_\lambda\rangle\le\|a\|\|Z_j-Z_\lambda\|\le L\|Z_j-Z_\lambda\|.
\]
Thus $\mathcal D_N$ is an $L$-risk field. For every $Y\in L^2_{\mathbb P}(\mathcal F)$, the objective defining $\rho_{\mathcal D_N}(Y)$ is real-valued and continuous on the nonempty compact simplex $\Delta_N$, so its minimum exists and is finite. Since $\mathbb E[Z_AY]=\mathbb E[Y]-\langle a,P_{\mathbf{1}}(Y)\rangle$, the reconstruction formula in Theorem~\ref{thrm:Reconstruction} gives
\begin{align*}
\rho_{\mathcal D_N}(Y)+\mathbb E[Z_AY]
&=\min_{\lambda\in\Delta_N}\left\{L\|P_{\mathbf{1}}(Y)-Z_\lambda\|-\langle a,P_{\mathbf{1}}(Y)-Z_\lambda\rangle\right\}\\
&\ge\min_{\lambda\in\Delta_N}\left\{(L-\|a\|)\|P_{\mathbf{1}}(Y)-Z_\lambda\|\right\}\ge0.
\end{align*}
For $X,Y\in L^2_{\mathbb P}(\mathcal F)$ with $Y\le X$ $\mathbb P$-a.s., it follows that $\rho_{\mathcal D_N}(Y)\ge-\mathbb E[Z_AY]=-\mathbb E[Z_AX]+\mathbb E[Z_A(X-Y)]\ge-\mathbb E[Z_AX]$. Taking the infimum over such $Y$ in~\eqref{eq:monotone_hull_reconstruction}, and using the admissible choice $Y=X$, yields: for each $X\in L^2_{\mathbb P}(\mathcal F)$
\begin{equation}
\label{eq:ES_common_tail_sandwich}
-\infty<-\mathbb E[Z_AX]\le\rho_{\mathcal D_N}^{\uparrow}(X)\le\rho_{\mathcal D_N}(X)<\infty.
\end{equation}
Fix $w\in\Delta_N$, $k\in\mathbb R$, and $H\in\{Y\in L^2_{\mathbb P}(\mathcal F):Y\ge0,\ \mathbf{1}_AY=0\ \mathbb P\text{-a.s.}\}$.
Set
$
 c \eqdef
 \sum_{n=1}^Nw_nc_n$, 
$V\eqdef\sum_{n=1}^Nw_nX_n+k\mathbf{1}$, $
X\eqdef V+H$; and since $H\ge0$ and $\mathbf{1}_AH=0$, we have
$\mathbb P$-a.s.
\begin{align*}
\mathbf{1}_A\bigl(X-( c +k)\mathbf{1}\bigr)
&=\sum_{n=1}^Nw_n\mathbf{1}_A
  (X_n-c_n\mathbf{1})\le0
,\\
\mathbf{1}_{A^c}\bigl(X-( c +k)\mathbf{1}\bigr)
&=\sum_{n=1}^Nw_n\mathbf{1}_{A^c}
  (X_n-c_n\mathbf{1})+H\ge0.
\end{align*}
Applying~\eqref{eq:ES_common_tail_linearization} with
$U=X$ and $d= c +k$, and using $Z_AH=0$, gives
\[
\operatorname{ES}_{\alpha}(-X)
=-\mathbb E[Z_AX]
=-\mathbb E[Z_AV]
=\sum_{n=1}^Nw_ny_n-k
\]
Since $V\le X$, the definition of the monotone hull,
\eqref{eq:ES_common_tail_sandwich}, and the reconstruction
formula evaluated at $\lambda=w$ give
\begin{align*}
\operatorname{ES}_{\alpha}(-X)
\le \rho_{\mathcal D_N}^{\uparrow}(X)
\le \rho_{\mathcal D_N}(V)
\le \sum_{n=1}^N w_ny_n-k
=\operatorname{ES}_{\alpha}(-X).
\end{align*}
Taking $k=0$ proves~\eqref{eq:ES_common_tail_exactness}.
\end{proof}

\section{Randomized  {Dual Representation}}
\label{s:ToyExperiments__ss:Computation}

 {Direct evaluation} of the reconstruction formula  {\eqref{eq:reconstruction_formula} requires solving the primal optimization problem
\[
    \phi(z)
    =
    \min_{\lambda\in\Delta_{N_{\mathrm{train}}}}
    \left\{
        \tilde y^\top\lambda
        +
        L\|D_p^{1/2}(z-Z\lambda)\|_2
    \right\},
\]
where $Z\in\mathbb{R}^{B\times N_{\mathrm{train}}}$ has columns
$
z_n=x_n-(p^\top x_n)\mathbf{1},
$
and $\tilde{y}$ has components
$
\tilde y_n=y_n+p^\top x_n.
$
The optimization variable is the simplex weight vector
$\lambda\in\Delta_{N_{\mathrm{train}}}$, so the problem dimension grows with the number of elicited quotes.} 
Instead, we exploit the  {Fenchel} dual representation
\begin{equation}
\label{eq:conv_dual}
    \phi(z)
    =
    \max_{u\in\mathbb{R}^B,\;t\in\mathbb{R}}
        \left\{u^\top D_p^{1/2}z+t\right\}
  \quad  \text{subject to}\quad
  \begin{cases}
        & t\le
        \tilde y_n-u^\top D_p^{1/2}z_n,
        \,
        n\in [N_{\mathrm{train}}]_+,\\
& \|u\|_2\le L.
        \end{cases}
\end{equation}
 {The dual contains only $B+1$ optimization variables, namely $(u,t)$, instead of $N_{\mathrm{train}}$ simplex variables. Consequently, whenever $B\ll N_{\mathrm{train}}$, it provides a substantially lower-dimensional representation.}
 {To further accelerate evaluation, we replace the dual problem by a randomized finite supporting-hyperplane approximation.} We sample $M_{\mathrm{dual}}$ directions $u_1,\ldots,u_{M_{\mathrm{dual}}}$ uniformly from the sphere $\{u\in\mathbb{R}^B:\|u\|_2=L\}$, and  {precompute the intercepts}
\[
     b_m
    \eqdef
    \min_{1\le n\le N_{\mathrm{train}}}
    \left\{
        \tilde y_n-u_m^\top D_p^{1/2}z_n
    \right\},\qquad
m=1,\ldots,M_{\mathrm{dual}}.
\]
 {This yields the approximation
\[
\phi_{M_{\mathrm{dual}}}(z)
=
\max_{1\le m\le M_{\mathrm{dual}}}
\left\{
u_m^\top D_p^{1/2}z+b_m
\right\},
\]
which is the maximum of finitely many affine functions. After the offline computation of the intercepts $b_m$, evaluating
$\phi_{M_{\mathrm{dual}}}$ requires only a matrix-vector multiplication followed by a maximum over $M_{\mathrm{dual}}$ affine functions. Since $\phi_{M_{\mathrm{dual}}}$ is the pointwise maximum of affine functions, it remains convex and $L$-Lipschitz.} Consequently, its cash-additive lift
\[
    \rho_{\mathcal D_N}^{(M_{\mathrm{dual}})}(x)
    \eqdef
    \phi_{M_{\mathrm{dual}}}\bigl(x-(p^\top x)\mathbf{1}\bigr)-p^\top x
\]
is again cash-additive by construction.  {Unless stated otherwise, all experiments use
$M_{\mathrm{dual}}=10^5$ sampled dual directions.}

\end{appendix}

\bibliographystyle{acm}
\bibliography{Bookkeaping/Refs}

\end{document}